\documentclass{ieeetmlcn}
\usepackage{cite}
\usepackage{amsmath,amssymb,amsfonts}
\usepackage{algorithmic}
\usepackage{graphicx,color}
\usepackage{textcomp}
\usepackage{xcolor}
\usepackage{hyperref}
\hypersetup{hidelinks} 
\usepackage{algorithm,algorithmic}
\def\BibTeX{{\rm B\kern-.05em{\sc i\kern-.025em b}\kern-.08em
    T\kern-.1667em\lower.7ex\hbox{E}\kern-.125emX}}
\AtBeginDocument{\definecolor{tmlcncolor}{cmyk}{0.93,0.59,0.15,0.02}\definecolor{NavyBlue}{RGB}{0,86,125}}
\usepackage{orcidlink}
\usepackage{multirow} 
\usepackage{multicol} 
\newcommand{\specialcellT}[2][c]{ 
  \begin{tabular}[#1]{@{}c@{}}#2\end{tabular}}

\def\OJlogo{\vspace{-4pt}\includegraphics[height=18pt]{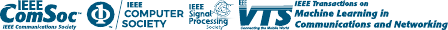}}
\def\seclogo{\vspace{10pt}\includegraphics[height=18pt]{new_logo}}

\def\authorrefmark#1{\ensuremath{^{\textbf{#1}}}}
\usepackage[normalem]{ulem} 
\usepackage{stfloats} 
\usepackage{placeins} 

\begin{document}
\receiveddate{19 October, 2023}
\reviseddate{10 August, 2024 and 5 December 2024}
\accepteddate{8 February, 2025}
\publisheddate{17 February, 2025}
\currentdate{20 February, 2025}
\doiinfo{TMLCN.2025.3542760}

\markboth{}{Author {et al.}}

\title{RACH Traffic Prediction in Massive Machine Type Communications}

\author{Hossein~Mehri\orcidlink{0000-0002-6949-9298}\authorrefmark{1} ~(Student~Member,~IEEE), Hani~Mehrpouyan\authorrefmark{2}~(Member,~IEEE),\\ and Hao~Chen~(Member,~IEEE)\orcidlink{0000-0001-6212-7718}\authorrefmark{1}}
\affil{Department of Electrical and Computer Engineering, Boise State University, Boise, ID 83725 USA}
\affil{Mehrs LLC, Island Park, ID 83429 USA}
\corresp{Corresponding author: H.~Mehri (email: hosseinmehri@u.boisestate.edu).}
\authornote{a.\copyright 2025 The Authors. This work is licensed under a Creative Commons Attribution 4.0 License.\\For more information, see https://creativecommons.org/licenses/by/4.0/}

\begin{abstract}
Traffic pattern prediction has emerged as a promising approach for efficiently managing and mitigating the impacts of event-driven bursty traffic in massive machine-type communication (mMTC) networks. However, achieving accurate predictions of bursty traffic remains a non-trivial task due to the inherent randomness of events, and these challenges intensify within live network environments. Consequently, there is a compelling imperative to design a lightweight and agile framework capable of assimilating continuously collected data from the network and accurately forecasting bursty traffic in mMTC networks. This paper addresses these challenges by presenting a machine learning-based framework tailored for forecasting bursty traffic in multi-channel slotted ALOHA networks.
The proposed machine learning network comprises long-term short-term memory (LSTM) and a DenseNet with feed-forward neural network (FFNN) layers, where the residual connections enhance the training ability of the machine learning network in capturing complicated patterns. Furthermore, we develop a new low-complexity online prediction algorithm that updates the states of the LSTM network by leveraging frequently collected data from the mMTC network.
Simulation results and complexity analysis demonstrate the superiority of our proposed algorithm in terms of both accuracy and complexity, making it well-suited for time-critical live scenarios. We evaluate the performance of the proposed framework in a network with a single base station and thousands of devices organized into groups with distinct traffic-generating characteristics. Comprehensive evaluations and simulations indicate that our proposed machine learning approach achieves a remarkable $52\%$ higher accuracy in long-term predictions compared to traditional methods, without imposing additional processing load on the system.
\end{abstract}



\maketitle

\section{INTRODUCTION}
\IEEEPARstart{T}{he} Internet of Things (IoT) has emerged as one of the most important technologies of the twenty-first century. In recent years, the number of new IoT products and applications has grown exponentially, and IoT technology has been extensively applied across diverse fields such as smart cities, smart transportation, smart factories, smart healthcare, intelligent warning systems, and disaster management~\cite{[ref1]}. Automated devices such as sensors, tracking devices, and meters, often referred to as machine-type communication (MTC) devices, are assumed to have a density of one million devices per square kilometer~\cite{[ref2],[ref3],[ref4]}. These networks can be characterized by a massive amount of data available through the devices and the possibility of exchange of this information among the large number of devices without or with little intervention by human~\cite{[ref1]}. Examples of MTC application in smart manufacturing and industrial applications are discussed in~\cite{[ref604]} and~\cite{[ref605]}, demonstrating significant improvements in operational efficiency through real-time data acquisition, fault detection, and intelligent decision-making. However, these advancements also result in increased traffic from numerous interconnected devices, highlighting the need for effective traffic management to prevent network congestion. Although significant progress has been made by the 3rd generation partnership project (3GPP) to make 5G mobile and wireless networks support MTC, it can still be a challenging task to transfer the large amount of data collected by massive number of devices using the existing infrastructures~\cite{[ref6]}.

With the popularity of MTC applications, significant research is underway to upgrade the existing algorithms that are adopted in long term evolution advanced (LTE-A) for MTC for use in 5G networks. These algorithms are categorized as either contention-based or contention-free methods. Among them, the contention-based multi-channel slotted ALOHA scheme is widely used in uplink random access channels (RACH)~\cite{[ref9],[ref11]}. This is due to the scheme's low a priori knowledge requirement, in which devices randomly select channels and transmit preambles/packets to the associated base station without negotiation~\cite{[ref12]}.
The contention-based approach is suitable for a network with limited number of access requests~\cite{[ref13]}. However, this scheme may fail in massive MTC (mMTC) networks during bursty traffic in which tens of thousands devices need to reserve the wireless access resources and transmit their packet in a short period of time~\cite{[ref13]}. Collisions are thus the primary issue with the contention-based slotted ALOHA schemes in mMTC networks~\cite{[ref10]}.

On the other hand, contention-free schemes suffer from high complexity and scalability issues. For example, in grant-free schemes~\cite{[ref17],[ref18]}, a user is immediately assigned a unique non-orthogonal preamble (similar to an identification number) upon joining the network and uses this preamble to transmit its packets to the network. Employing no-orthogonal preambles resolves the collision issue in the network at the cost of increased user detection complexity at the receiver. Moreover, the total number of devices that can exist in a network is limited by the available non-orthogonal preambles~\cite{[ref19]}.
In this work, we focus on designing a method to address the congestion issue in contention-based multi-channel ALOHA schemes.

\subsection{Motivation}
The traffic pattern of MTC networks can generally be categorized into periodic and event-driven. Periodic traffic is characterized by a regular but sporadic pattern and accounts for the majority of traffic in mMTC networks~\cite{[ref20]}. This type of traffic occurs when devices communicate with the base station on a regular basis to provide status reports and updates. Due to sparsity of packet generation in this type of traffic, existing network resources will be sufficient for the massive number of devices in the network~\cite{[ref20]}. Smart meter readings are a typical example of this form of traffic, e.g., gas, electricity, water. The event-driven traffic is generated as a reaction to some event and typically results in a burst of network access requests. 3GPP technical reports identifies two types of events that triggers a bursty traffic in a network~\cite{[ref21],[ref22]}: 1) When many MTC devices are activated by the MTC server at the same time; 2) a large number of MTC devices send connection/activation/modify requests at the same time due to an external event. Unlike the periodic traffic, event-driven bursty traffic can cause a large number of collisions in the network~\cite{[ref20],[ref21]}. A large number of access requests after a power outage is an example of event-driven traffic.

Limited number of access channels and the randomness of choosing these channels by users are the main causes of collisions in the network \cite{[ref23]}. 5G new radio (NR) and LTE-A standards both use a similar procedure for random access, in which a set of orthogonal preambles are used in initial access procedure~\cite{[ref24],[ref23]}. Each device with a packet to transmit, randomly chooses a preamble and sends its request to the base station. When two or more devices choose the same preamble for their transmission, a collision occurs and all packets are lost. As the number of requests on the network increases, the probability of collision increases as well. As the number of available orthogonal preambles is limited, certain procedures are required to manage the high number of collisions that occur during bursty traffic~\cite{[ref25],[ref26]}. The majority of existing procedures operate in reactive mode, i.e., collisions are handled as congestion occur~\cite{[ref25]}. On the other hand, proactive methods can initiate the traffic management procedures prior to the occurrence of traffic congestion in the network~\cite{[ref23],[ref25]}. These methods require precise predictions of traffic for the upcoming time slots~\cite{[ref27]}. The demand for accurate prediction of the network traffic and the probability of occurrence of congestion a few seconds ahead of time motivates the authors of this paper to propose a new machine learning (ML) based prediction method. 

The random and non-linear behavior of bursty traffic~\cite{[ref28],[ref29]} makes the mMTC traffic pattern a non-stationary and non-linear time series that is hard to model using traditional modeling methods~\cite{[ref12],[ref30]}. On the other hand, ML-based methods have demonstrated superior performance, particularly in addressing highly complex problems (\hspace{-1sp}\cite{[ref12]}), and have been widely applied to time-series prediction problems~\cite{[ref31], [ref32], [ref33]}. Among these, long-term short-term memory (LSTM) networks have exhibited exceptional capability in extracting temporal features from sequential patterns and have been extensively used in traffic prediction studies~\cite{[ref33]}.

In the context of time-series forecasting with LSTM networks, two distinct scenarios are commonly encountered: offline and live scenarios. In offline scenarios, observed samples (historical data) are utilized to initialize the states of the LSTM network, and then make predictions for the next few steps. However, in live scenarios, the availability of fresh data at frequent intervals necessitates a different approach.
To handle live scenarios, a method known as the rolling algorithm is often employed. This algorithm follows a similar strategy to the offline method, where fresh data is concatenated with historical data, and the oldest samples are removed. Subsequently, the updated set of data points is fed into the LSTM network to generate predictions for future time steps. While the rolling algorithm can generate continuous predictions in live scenarios, it comes with certain inefficiencies. Notably, it necessitates data buffering to manage historical information effectively. Additionally, the concatenation and removal of samples at each step can introduce computational overhead, affecting the prediction speed and overall efficiency of the method.
To address this issues, we propose a novel online prediction algorithm which differs from the rolling algorithm in that it solely utilizes fresh data to update the LSTM states, eliminating the need for data buffering and avoiding computational redundancies associated with the rolling algorithm. As a result, the proposed online algorithm offers several advantages over the rolling method, including increased speed, improved accuracy, and no data buffering requirements. This makes the proposed algorithm well-suited for addressing the mMTC network traffic prediction problem, where continuous monitoring and prompt response to events are crucial for effective network maintenance.

Furthermore, despite the existing online learning methods where fresh data is used to retrain the LSTM network and update the LSTM parameters, the proposed online prediction algorithm focuses on the prediction phase, where only network states are updated. Moreover, online learning algorithms are particularly suitable for scenarios in which the statistics of the data under investigation are changing dynamically, necessitating parameter updates, that is not the case in this paper. Consequently, in this work, fresh data is exclusively employed to update the LSTM states, thereby circumventing the challenges associated with online learning methods, such as \textit{catastrophic forgetting} and \textit{catastrophic interference}~\cite{[ref37],[ref38]}. By adopting this approach, potential issues of the network's inability to retain previously learned information or interference with existing knowledge during parameter updates are effectively mitigated.

\subsection{Prior Works}
Analytical models to investigate the transient behavior of the RACH channel in orthogonal frequency division multiple access (OFDMA) system under bursty traffics are studied in~\cite{[ref39]}. Their analysis shows that bursty traffic may cause congestion on the RACH channel, resulting in intolerable delays, packet loss, or even service unavailability in the network. 
3GPP established multiple congestion avoidance mechanisms to deal with the massive number of simultaneous IoT connection requests in RACH~\cite{[ref40],[ref41],[ref13]}. The first mechanism is access class barring (ACB), defined by 3GPP in Release 8, and the second is extended access barring (EAB), defined in Release 11, both with the goal of preventing overload of the access network~\cite{[ref28]}. Many implementations of ACB and EAB have been introduced in the literature to avoid congestion at the access network~\cite{[ref42],[ref43],[ref44]}. Although these implementations typically require a precise estimate of traffic status in the following time slots, they used a simplified model that did not account for bursty events, which may fail in practical situations~\cite{[ref45]}. 

The use of ML-based methods to predict network traffic and to alleviate congestion has recently received extensive attention. 
The authors of~\cite{[ref31]} and~\cite{[ref49]} used LSTM network to predict the source traffic of devices during bursty traffic in MTC networks. The predicted traffic of individual devices (source traffic) can be used by resource allocation algorithms, e.g., fast uplink grant schemes~\cite{[ref52],[ref50],[ref51]}, to efficiently utilize the wireless resources and improve the service quality~\cite{[ref53]}. Nonetheless, the main drawback of source traffic prediction methods is scalability, especially when dealing with mMTC networks with a large number of devices. In this work, instead of predicting the traffic of each individual device, we focus on the overall traffic of the network in order to prevent network congestion.

Most of the existing methods of predicting the traffic pattern focus on human type traffic or ignores the event-driven bursty traffic~\cite{[ref31],[ref49],[ref52],[ref54],[ref55],[ref40]}. 
A Q-learning-based method is used in~\cite{[ref54]} to collaboratively avoid congestion in MTC networks. However, this method suffers from the scalability and can only be applied to non-bursty conditions where the number of active devices is limited. Duc-Dung \textit{et al.} \cite{[ref55]} adopted a Q-learning-based method to assign devices to random slots in non-orthogonal multiple access (NOMA) based mMTC systems. A hybrid ACB/EAB-based congestion avoidance method is proposed in~\cite{[ref40]} that changes the policy of access to the network based on the probability of bursty congestion in future time slots. Due to lack of capable bursty event prediction mechanisms in these works (\hspace{-1sp}\cite{[ref31],[ref49],[ref52],[ref54],[ref55],[ref40]}), they typically ignore the bursty events or use simple models to represent the bursty traffic in their work, which significantly affects their performance in real-world applications.

In~\cite{[ref23]} and~\cite{[ref57]}, the authors proposed an ML based method to predict the MTC network traffic during the bursty scenarios. Their work focuses exclusively on predicting traffic pattern after observing an event and without considering the uniform traffic of the devices. However, bursty and uniform traffic behave differently, and while neither alone may cause congestion, the combination of the two can. This necessitates the development of a ML network capable of predicting both uniform and bursty traffic in real-world scenarios. Table \ref{table-1} summarizes the related works on traffic prediction in MTC networks.
\begin{table*}[htbp]
\caption{Summary of Related Work on Traffic Prediction in mMTC Networks.}
\label{table-1}
\centering
\begin{tabular}{|c|c|c|c|c|c|}
\hline
 \textbf{Reference} & \textbf{Main Focus} & \textbf{Prediction Model} & \textbf{Target Traffic Type} & \textbf{MTC Groups} & \textbf{Online Prediction} \\ \hline 
 \cite{[ref23]} & Traffic prediction & LSTM & Aggregated traffic & Single & \texttimes \\ \hline
 \cite{[ref31]} & \specialcellT{{Traffic prediction,}\\ {resource allocation}} & Direct information & Source traffic & Single & \texttimes \\ \hline
 \cite{[ref43]} & Resource allocation & \specialcellT{{Statistical models}\\ {(fluid model extension,}\\ {crowd-source information,}\\ {recursive estimation})} & Aggregated traffic & Single & \texttimes \\ \hline
 \cite{[ref49]} & \specialcellT{{Traffic prediction,}\\ {resource allocation}} & LSTM & Source traffic & Single & \texttimes \\ \hline
 \cite{[ref50]} & Resource allocation & LSTM, support vector machine (SVM) & Source traffic & Single & \texttimes \\ \hline
 \cite{[ref51]} & Resource allocation & Statistical models & Source traffic & Single & \texttimes \\ \hline
 \cite{[ref52]} & Resource allocation & On-Off Markov process & Source traffic & Single & \texttimes \\ \hline
 \cite{[ref53]} & Resource allocation & LSTM, SVM, KNN & Aggregated traffic & Single & \texttimes \\ \hline
 \cite{[ref57]} & Traffic prediction & LSTM & Aggregated traffic & Single & \texttimes \\ \hline
 This work & Traffic prediction & LSTM, GRU & Aggregated traffic & Multiple & \checkmark \\ \hline
\end{tabular}
\end{table*}

\subsection{Traditional Rolling Algorithm}

Although LSTM model outperforms typical recurrent neural networks (RNNs), it cannot maintain its accuracy over a longer period of time in the multi-step prediction of the time-series~\cite{[ref33]}. Consequently, prediction length in offline methods are limited to a few steps based of the the time-series properties. On the other hand, longer and continuous predictions are possible in live scenarios where fresh data is available every often. In live scenarios, the traditional rolling method is often utilized for live predictions. This method involves frequent updates to historical data by concatenating new data while removing the oldest samples. Subsequently, similar to offline methods, the LSTM network is initialized with the updated historical data, enabling predictions for the next few steps. The drawbacks of this method include the need for a large buffer to store historical data, extra computational overhead due to updating the historical data, and the time-consuming nature of initializing the LSTM with extended historical data. These factors make this method undesirable for time-sensitive and resource-limited applications. Moreover, a smaller buffer size deprives the longer dependencies from the time-series sequence and thereby reduces prediction accuracy. To address these challenges, we propose a novel algorithm that replaces the traditional rolling method, offering a lighter and faster approach for live predictions. The idea behind this approach is that all the extracted information from the historical data is abstracted in hidden states of LSTM network. Thus, instead of extracting the dependencies from the updated historical data upon arrival of fresh data, we restore the LSTM states and update them with the fresh data. This makes the initialization process much faster than rolling method.

\subsection{Contributions}
In this work, we present a machine learning based framework to predict the successfully detected preambles in uplink random access channel and congested bursty events in mMTC networks. This information can be used by network controlling agents to avoid network congestion. The contributions of this paper can be summarized as follows:
\begin{itemize}
    \item We adopt a machine learning (ML) network architecture that leverages Long Short-Term Memory (LSTM) and DenseNet models, resulting in enhanced learning ability and prediction accuracy without compromising training complexity.
    \item We present the Fast LiveStream Predictor (FLSP) algorithm, an innovative and efficient prediction method designed to replace traditional rolling algorithms in live scenarios. This novel approach leverages the frequently collected data from the MTC network to correct itself and improve prediction accuracy by maintaining the longer dependencies between time-series samples. Moreover, FLSP is applicable to any live time-series prediction problem as well as practical applications that require continuous and precise predictions.
    \item The complexity of FLSP algorithm is carried out and compared to the existing algorithm, demonstrating a significant improvement over the rolling method in terms of computational efficiency.
    \item To complement the traffic prediction network, we design an auxiliary ML network which receives the outputs of the traffic prediction network and accurately predicts congestion occurrences in the future.
    \item In order to evaluate the performance of the proposed method, we created a more realistic traffic pattern, encompassing both uniform and event-driven traffic. This traffic is generated by several groups of devices with different sizes and probabilities of event occurrence, effectively simulating real-world conditions.
\end{itemize}
By introducing FLSP algorithm, conducting comprehensive complexity analysis, and employing realistic traffic patterns for evaluation, our work significantly advances the state-of-the-art in traffic pattern prediction for mMTC networks. The proposed method not only enhances prediction accuracy and computational efficiency but also demonstrates broad applicability for real-world scenarios, further contributing to the field of time-series prediction and network management in mMTC environments.
\subsection{Acronyms}
To improve readability, all acronyms used in this paper are defined in Table~\ref{table0}. The acronyms are sorted in alphabetical order for ease of reference.
\begin{table}[htbp]
\caption{List of Acronyms (sorted alphabetically)}
\label{table0}
\centering
\begin{tabular}{|c|l|}
\hline
\textbf{Acronym} & \textbf{Definition} \\ \hline
3GPP & 3rd generation partnership project  \\ \hline
5G & Fifth generation of cellular technology  \\ \hline
ACB & Access class barring  \\ \hline
AOI & Age of information  \\ \hline
CNN-1D & One-dimensional convolutional neural network \\ \hline
EAB & Extended access barring  \\ \hline
FFNN & Feed-forward neural network  \\ \hline
FLOP & Floating-point operation \\ \hline
FLSP & Fast LiveStream Predictor  \\ \hline
GRU & Gated recurrent unit \\ \hline
IoT & Internet of Things  \\ \hline
LSTM & Long-term short-term memory  \\ \hline
LTE-A & Long term evolution advanced  \\ \hline
ML & Machine learning  \\ \hline
mMTC & Massive machine-type communication  \\ \hline
MSE & Mean square error  \\ \hline
MTC & Machine-type communication  \\ \hline
NOMA & Non-orthogonal multiple access  \\ \hline
NR & New radio  \\ \hline
OFDMA & Orthogonal frequency division multiple access  \\ \hline
RACH & Random access channels  \\ \hline
RNN & Recurrent neural network  \\ \hline
\end{tabular}
\end{table}

\subsection{Organization}
This paper is organized as follows. In section~\ref{SysModel}, the system model and definitions are specified. The ML networks for traffic pattern and congestion prediction are described in section~\ref{MLNet}. The FLSP algorithm is explained in details in section~\ref{OnlineMethod}. Training procedure of the proposed burst detection network is described in section~\ref{Burst Detection}. Simulation results are presented and discussed in section~\ref{Sim}. Finally, conclusion is provided in section~\ref{Conc}.

\section{System Model and Definitions}
\label{SysModel}

\begin{figure*}[htbp]
\centerline{\includegraphics[scale=0.5]{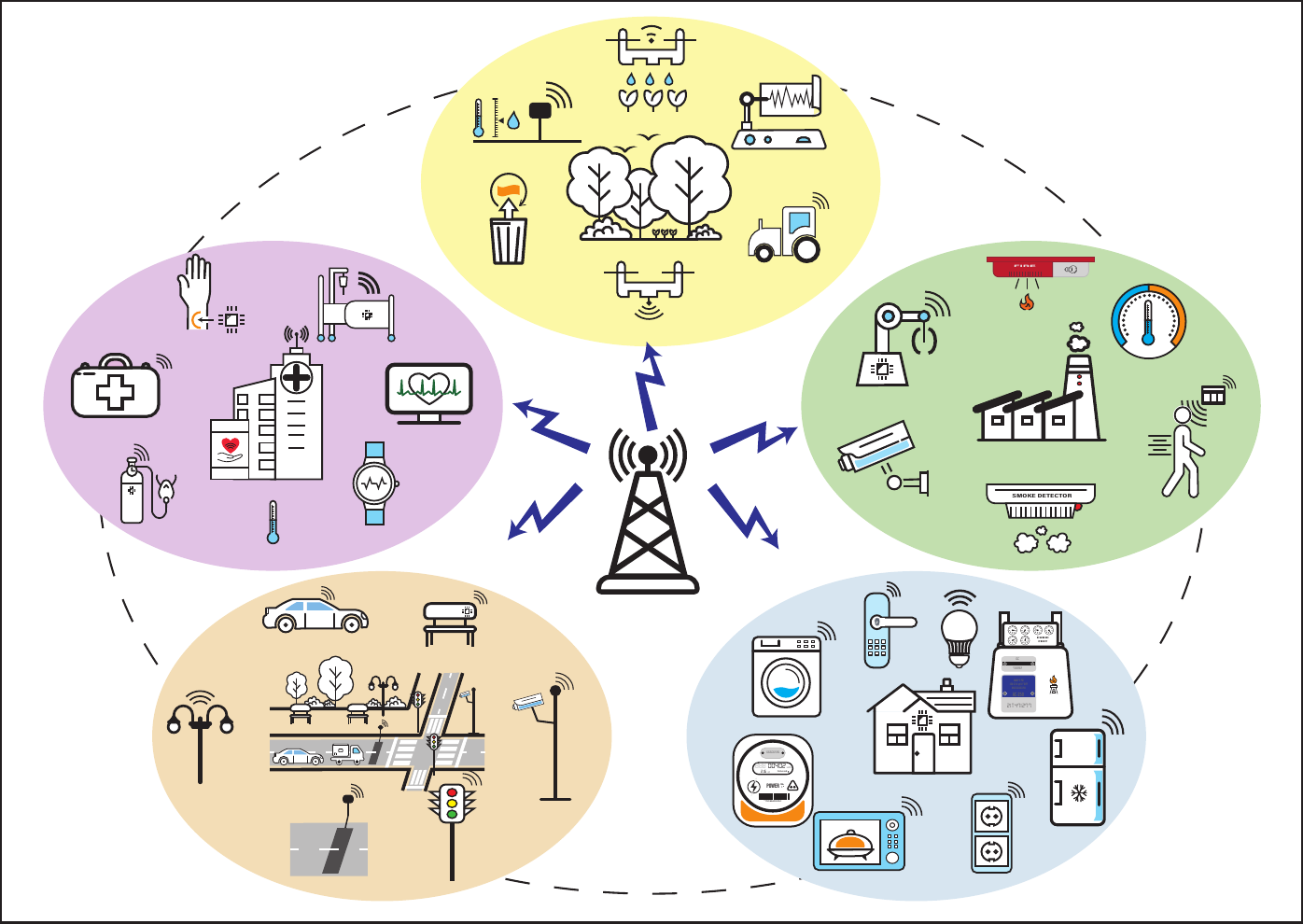}}
\caption{An example of application-based categorization of devices in an mMTC network. The total network traffic is the sum of the traffic generated by the various groups, each with their own set of characteristics.}
\label{FigureSM}
\end{figure*}
In this paper, we consider a large number of devices that are organized into several groups. This grouping can be based on the application, functionality, or geographical location of the devices. Fig.~\ref{FigureSM} depicts an example of application-based grouping. In general, the size and traffic characteristics of these groups may vary from one to the next. The following subsections describe the channel model, definitions, and traffic models that are considered through out this paper.

\subsection{Channel Description and Transmission Properties}

\begin{figure}[htbp]
\centerline{\includegraphics[scale=0.6]{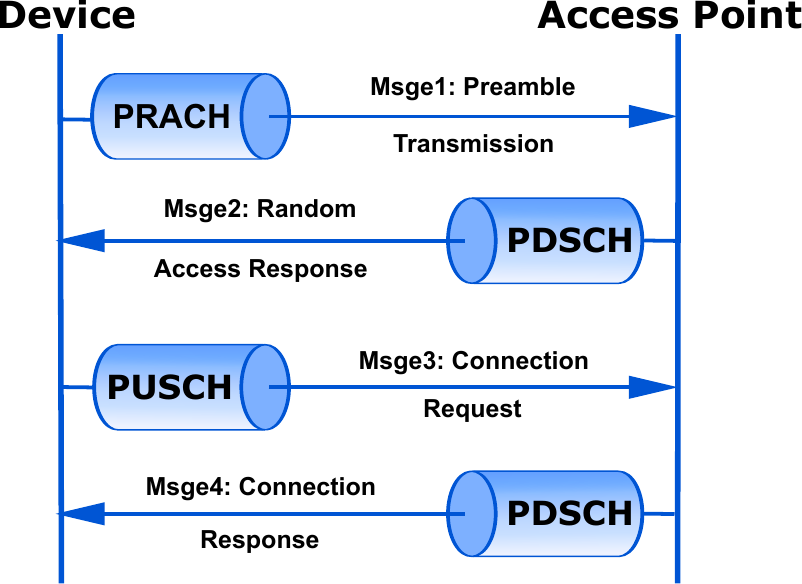}}
\caption{Four-way handshake procedure of access request in LTE and NR-5G.}
\label{figure1}
\end{figure}
The random access procedure normally involves a four-message handshake between a device and the base station (Fig.~\ref{figure1}). The four messages include a Preamble Transmission (Message 1) initiated by a device or user; a Random Access Response (Message 2) sent in response by the base station; a Connection Request (Message 3) transmitted by the device or user ; and a Connection Response (Message 4) confirmed by the base station~\cite{[ref58],[ref39]}. The first message is transmitted over a contention-based access medium, i.e., the RACH channel, through which many devices are attempting to send their access request to the network. Hence, successfully transmitting Message 1 over the RACH can be a limiting factor since, the RACH is open to collision and congestion. The well-known ALOHA scheme is used over the RACH in which each device randomly selects a preamble and transmits its request in the first available random-access slot. The purpose of this paper is to develop a framework for predicting successfully transmitted and congested preambles in the network.

\subsection{Definitions}
\label{Defs}
Below definitions are used through out of the paper:
\begin{itemize}
    \item \textit{Arrivals}: The number of devices that are ready to send a packet on the uplink path.
    \item \textit{Attempts}: Total number of transmissions at each time slot. Attempts equals arrivals plus the number of retransmissions due to collisions.
    \item \textit{Detected attempts}: Attempts that are successfully detected by base station.
    \item \textit{Unsuccessful transmissions}: Unsuccessful transmission only happens when two or more devices try to access the network using the same preamble. Otherwise, transmission will be successful with the probability of one.
    \item \textit{Collision}: Collision occurs when two or more devices try to access the network using the same preamble.
    \item \textit{Congestion}: In this paper, congestion is defined as a situation in which the number of detected attempts is less than the number of arrivals over a given time period. Unlike collision, which is defined for a preamble at each time slot, congestion is defined for a period of time during which the majority of attempts fail, resulting in packet drop after definite retransmissions.
\end{itemize}

\subsection{mMTC Traffic Models}
In this paper we considered a single mMTC cell with $N$ IoT devices which are divided into several non-overlapping groups with different sizes and traffic generating properties. The set of IoT devices is denoted by $G \triangleq \{g_l|l=1,2,...,L\}$, where $L$ denotes the total number of groups in the network. Each group of devices generates two types of data packets: periodic and event-driven data packets. The generation of periodic data packets (which typically contains regular environmental sensing data) follows a uniform distribution with the probability of $p_u$ packets per second. 

According to the 3GPP specifications, the event-driven traffic follows the Beta distribution~\cite{[ref28]}. Assuming that an event observation happens between $t=0$ and $t=T$ in group $g_l$ with $n_l$ devices, the random access intensity is described by the Beta distribution of $p(t)$. Subsequently, the expected number of devices with a new packet in the $i$-th time slot (arrivals) can be calculated as follows
\begin{equation}
    Arrivals(i)=n_l\int_{t_{i}}^{t_{i+1}}p(t)d(t),
    \label{eq1}
\end{equation}
where $t_i$ is the time of the $i$-th time slot. $p(t)$ can be calculated from Beta distribution:
\begin{equation}
    p(t)=\frac{t^{\alpha-1}(T-t)^{\beta-1}}{T^{\alpha+\beta-1}B(\alpha,\beta)},
    \label{eq2}
\end{equation}
where $B(\cdot)$ is the Beta function with values of $\alpha=3$ and $\beta=4$~\cite{[ref28]}.

\section{Proposed Architecture}
\label{MLNet}
In this section, we explain the proposed ML architecture and provide a comprehensive description of the Fast LiveStream Predictor (FLSP) algorithm, a novel approach designed to replace traditional rolling algorithm in time-series prediction problem in live scenarios. We also provide the complexity analysis of the proposed algorithm and compare it to the rolling algorithm.

\subsection{Proposed ML Architecture}
\begin{figure}[htbp]
\centerline{\includegraphics[scale=0.4]{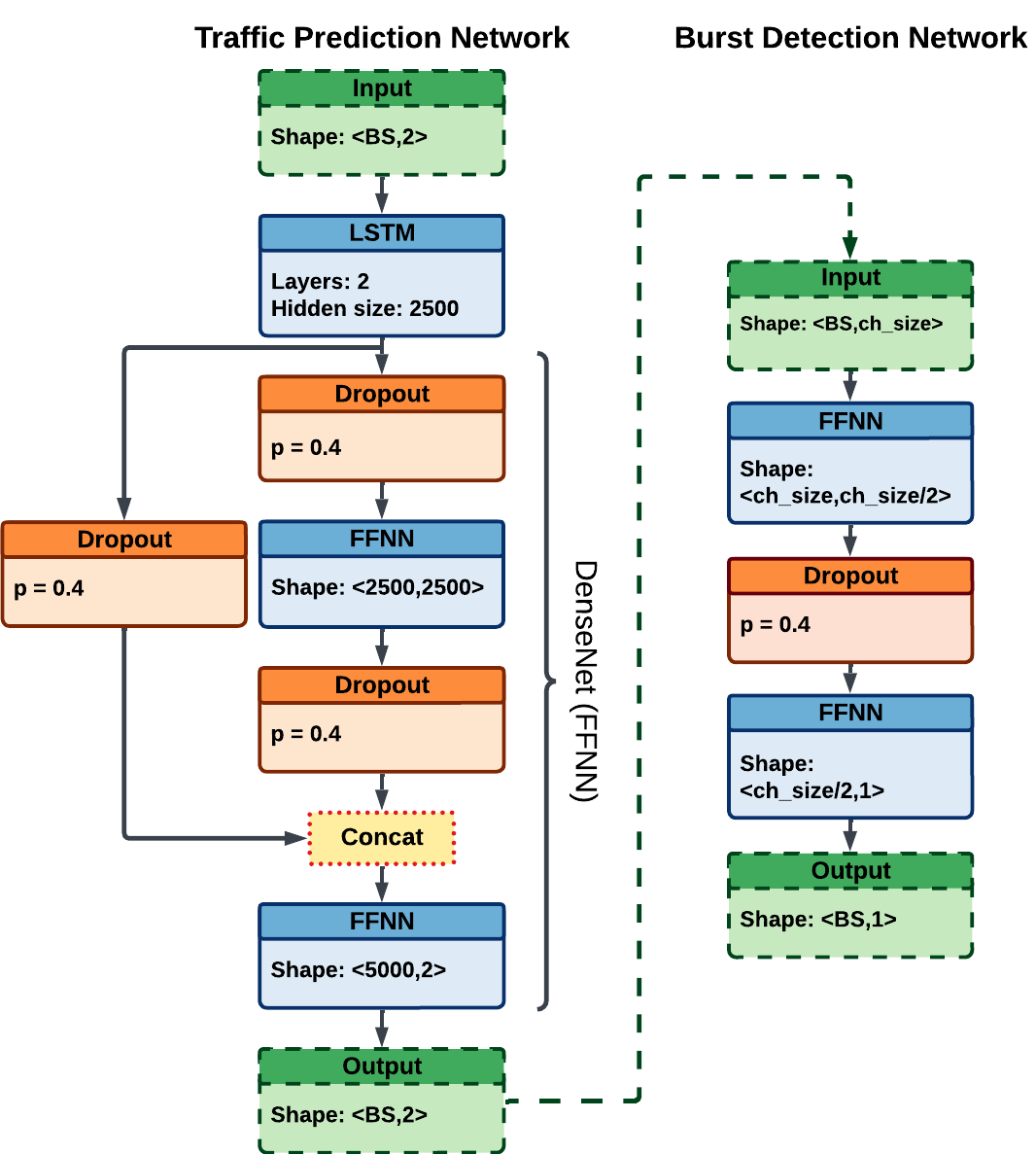}}
\caption{Proposed ML architectures for traffic prediction and congested bursty event detection networks. Second network receives the output of the first network as chunks of data.}
\label{figure2}
\end{figure}

Figure.~\ref{figure2} depicts the two main networks in our proposed ML architecture. The first network comprises a two-layer LSTM and a two-layer DenseNet with dropout layers introduced between the LSTM and DenseNet sections, as well as between the connections of the DenseNet layer, to mitigate overfitting during training. The combination of DenseNet and LSTM ensures a high convergence rate and effective extraction of both short-term and long-term dependencies in the traffic pattern.

The proposed ML architecture leverages the strengths of LSTM and DenseNet networks to accurately predict traffic pattern in mMTC networks. LSTM, a variant of RNN, has demonstrated superiority over traditional time series prediction methods~\cite{[ref32]}, overcoming the short-term memory constraints inherent in typical RNNs due to its gated architecture. ResNet~\cite{[ref59]} and DenseNet~\cite{[ref59-2]} are two network architectures that employ residual connections with identical weights between layers to mitigate the gradient vanishing/exploding problem in deep ML networks during training. These connections facilitate network optimization and training without introducing additional complexity to the system. ResNet utilizes feature summation for shortcuts, whereas DenseNet employs feature concatenation. Notably, DenseNet has demonstrated superior feature utilization efficiency, outperforming ResNet; however, with a larger number of parameters and requiring more memory due to feature concatenation~\cite{[ref59-3]}. In this study, we opt for DenseNet for the linear layers following the LSTM network, prioritizing accurate predictions.

The second network, the burst detection network, receives data chunks from the first network and predicts whether bursty congestion will occur in the following slots. This network consists of two FFNN layers with a dropout layer between them to prevent overfitting. The size of the first layer is determined by the size of the output data chunks from the first network.

In summary, the proposed ML architecture demonstrates a significant capabilities in predicting traffic patterns and congestion probabilities, making it a promising approach for efficient traffic management in mMTC networks.

\subsection{Proposed Algorithm for Online Mode Prediction}
\label{OnlineMethod}
After training the LSTM network with the training data, the FLSP algorithm is employed in the prediction phase to forecast the traffic volume in the upcoming time slots. The first step in the FLSP algorithm is to initialize the LSTM network by feeding it with the most recent historical data. Assuming that $f(\cdot)$ denotes the functionality of the LSTM network, the initialization procedure can be written as follows:
\begin{equation}
    \hat{y}_i=f(y_{i-1},C_{i-1},H_{i-1}),
    \label{eq33}
\end{equation}
where $\hat{y}_i$ presents the output at the $i$-th step, while $y_{i-1}$, $C_{i-1}$, and $H_{i-1}$  denote the actual input sequence, cell state, and hidden state, respectively, at the $(i-1)$-th step.

\begin{figure}[htbp]
\centerline{\includegraphics[scale=0.45]{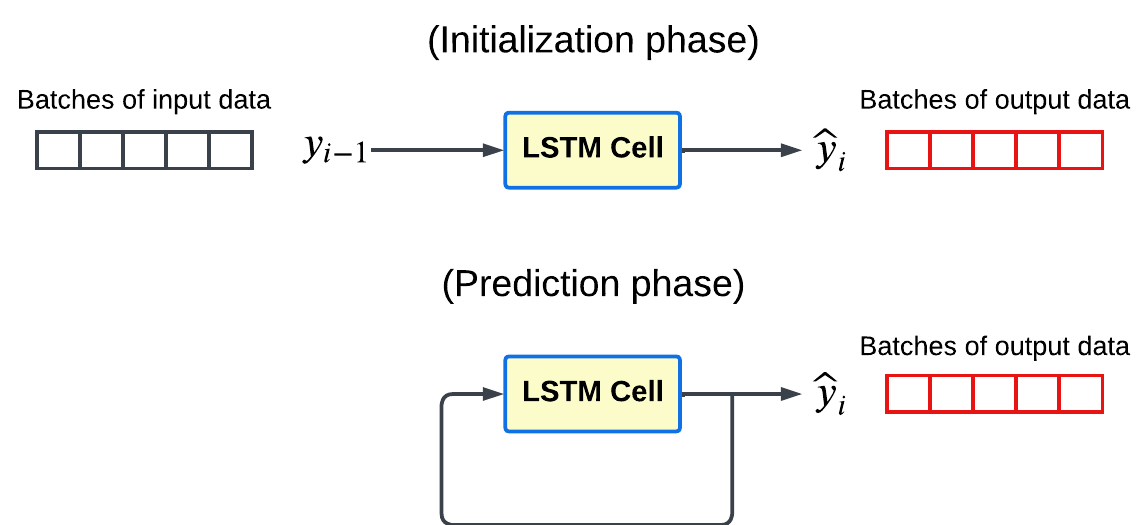}}
\caption{Recursive algorithm in the prediction phase of LSTM networks.}
\label{figure3}
\end{figure}

After initialization, the LSTM network is ready to generate predictions. A recursive algorithm is employed in the forward direction to predict sequential patterns. Based on this algorithm, the network's output from previous steps is used as input to predict the output at the current step. The recursive algorithm is depicted in Figure.~\ref{figure3} and is explained by the following equation:
\begin{equation}
    \hat{y}_{i+1}=f(\hat{y}_{i},C_{i},H_{i}),
    \label{eq44}
\end{equation}

The recursive algorithm is a fundamental component used in both offline and online methods, including both the rolling and proposed FLSP method. However, the imprecision of estimated values may lead to error propagation, resulting in deviations from the target values~\cite{[ref33]}. Moreover, LSTM states may no longer accurately represent the actual state of the ML network, which limits the length of predictions within the recursive algorithm. In live scenarios, on the other hand, the accuracy of predictions can be maintained by transforming the long-term prediction problem into several short-term prediction problems. In this case, upon the arrival of fresh data, predictions are made for a few steps ahead, and then the algorithm enters a standby mode, awaiting the receipt of new data points to correct itself and continue the prediction process. 

The FLSP algorithm leverages the fresh data to update the LSTM states and then make predictions. This algorithm stores the states of LSTM network before employing the recursive algorithm. These stored states of the LSTM network contain all the information from the historical data used for initialization and remain unaffected by the estimated values. When fresh data arrives, the FLSP algorithm restores these stored states and update them by incorporating the fresh data. This process is analogous to attaching the fresh data to the historical data and re-initializing the LSTM network, but is much faster. After updating the LSTM states with the fresh data and storing them, the recursive algorithm is employed to generate predictions. This procedure is repeated upon the arrival of fresh data and is described in greater detail in Algorithm~\ref{alg1}. 
\renewcommand{\algorithmiccomment}[1]{#1}
\begin{algorithm}
\caption{Fast LiveStream Prediction (FLSP)}
\label{alg1}
\begin{algorithmic}[1]
\STATE Get the $Hist$ data with the length of $l_{Hist}$.
\WHILE{$i<=l_{Hist}$}
    \STATE calculate $\widehat{y}_i=f(Hist_i)$
\ENDWHILE
\STATE Store the network states.
\WHILE{$i<=l_p$} 
    \STATE calculate $\widehat{y}_{l_{Hist}+i}=f(\widehat{y}_{l_{Hist}+i-1})$
\ENDWHILE
\STATE Wait to get $Fresh$ data with the length of $l_f$. \label{here3}
\STATE Restore the network states.
\WHILE{$i<=l_f$} 
    \STATE calculate $\overline{y}_i=f(Fresh_i)$
\ENDWHILE
\STATE Store network states.
\WHILE{$i<=l_p$} 
    \STATE calculate $\overline{y}_{l_f+i}=f(\overline{y}_{l_f+i-1})$
\ENDWHILE
\STATE Add only last $l_f$ steps of $\overline{y}$ to $\widehat{y}$.
\STATE Return $\widehat{y}$ and go to step~\ref{here3}.
\end{algorithmic}
\end{algorithm}

The primary distinction between the FLSP algorithm and the rolling algorithm lies in how they assimilate fresh data into their calculations. The FLSP algorithm utilizes fresh data to update the LSTM states, whereas the rolling algorithm attaches the fresh data to the historical data and then re-initialize the LSTM network with the updated historical data. This makes a significant difference in both computational requirements and memory usage to buffer the historical data. Moreover, the new method retains longer dependencies that might be lost in the traditional rolling method due to limited buffer size.

\begin{figure}[tbp]
\centerline{\includegraphics[scale=0.5]{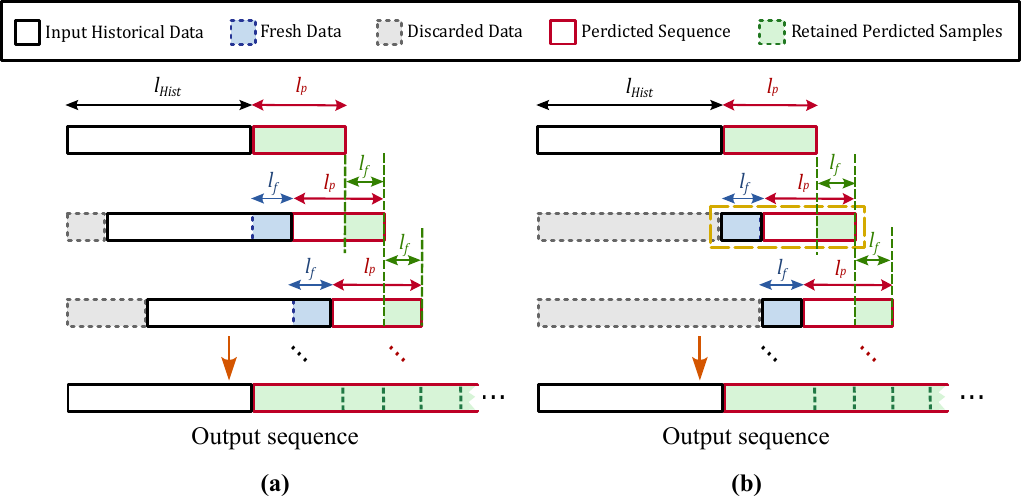}}
\caption{Comparison of rolling and FLSP algorithms in generating output sequences. (a) Rolling algorithm: The input sequence at each step (solid black frame) consists of truncated historical data and fresh data (blue shade). The traffic prediction network generates predictions for $l_p$ time slots at each step (solid red frame), but only the last $l_f$ time slots (green shade) are retained and appended to the output sequence. (b) FLSP algorithm: In this method, the input sequence includes only the fresh data, resulting in faster state updates compared to the rolling method. Yellow dashed lines represent a chunk of data, which is the concatenation of fresh data and predicted samples at each step, used as input for the burst detection network.}
\label{figure5new}
\end{figure}

Fig.~\ref{figure5new} illustrates an exemplary implementation of the rolling method and FLSP algorithm. In both methods, the LSTM network takes an input sequence, depicted by the solid black frame at each step. Notably, the FLSP algorithm uses significantly shorter input data for the LSTM network compared to the rolling method, resulting in reduced processing time. At each step of both methods, predictions are generated for a specific number of steps into the future. However, only the most recent steps are retained and appended to the output sequence. Assume that the network predicts $l_p$ steps ahead using the recursive method at each step after feeding the actual data. Then, the system remains in a state of readiness to receive the most recent data of size $l_f$, collected from the mMTC network (fresh data), in a way that $l_f<l_p$. After updating the LSTM states with the fresh data, the recursive algorithm generates the predictions for the next $l_p$ steps while retaining only the last $l_f$ steps of the predicted data. The reason for keeping only the last $l_f$ steps is that we are primarily interested in the unseen information, specifically the time slots that have not been predicted before, so we ignore the rest of the points when generating the output sequence. For example, when $l_p=200$ ($T_{pred}=1 [sec]$), $l_f=100$ ($T_{fresh}=0.5 [sec]$), and $l_{buff}=200$, the FLSP algorithm only receives fresh data of size $l_f=100$ time slots as input to the prediction model. In contrast, the rolling algorithm requires $l_{buff}+l_f=300$ time slots of data as input, which includes both buffered and fresh data. Both algorithms generate output for $l_p=200$ time slots, but only the last $l_f=100$ time slots are retained and appended to the output sequence. This results in an output sequence that aligns with the target sequenced used to calculate the evaluation metrics. The target sequence is the shifted version of the input sequence by $l_p$ time slots, representing the expected output from the prediction model at each time (ground truth data). It is important to note that the determination of prediction parameters depends on network configurations, the frequency of collecting fresh data, the required forecast length, and the desired prediction accuracy.

\subsection{Time Complexity}
To determine the time complexity of the proposed approach, we begin with an analysis of the time complexity of the multi-layer LSTM-DenseNet network during the forward pass. The following equations depict the forward pass of an LSTM cell:
\begin{equation}
\label{eq8_2}
\begin{split}
    i_{t} &=\sigma\left(W_{i i} x_{t}+b_{i i}+W_{h i} h_{t-1}+b_{h i}\right) \\
    f_{t} &=\sigma\left(W_{i f} x_{t}+b_{i f}+W_{h f} h_{t-1}+b_{h f}\right) \\
    g_{t} &=\tanh \left(W_{i g} x_{t}+b_{i g}+W_{h g} h_{t-1}+b_{h g}\right) \\
    o_{t} &=\sigma\left(W_{i o} x_{t}+b_{i o}+W_{h o} h_{t-1}+b_{h o}\right) \\
    c_{t} &=f_{t} \odot c_{t-1}+i_{t} \odot g_{t} \\
    h_{t} &=o_{t} \odot \tanh \left(c_{t}\right)
\end{split}
\end{equation}
The time complexity of all gates are similar and is in $\mathcal{O}(h(4d+4h+16))$ where $h$ denotes the size of the hidden layer of the LSTM, $d$ represents the feature size of the input sequence. Updating the LSTM states ($c_{t}$ and $h_{t}$) adds $\mathcal{O}(3h+2h)$ to the total time complexity of LSTM network. Thus, the time complexity of a multi-layer LSTM network with $L_{lstm}$ layers is in $\mathcal{O}(L_{lstm}h(4d+4h+21))$, or, after removing constants and simplifying, is in $\mathcal{O}(L_{lstm}hN)$, where $N=max(h,d)$.

The time complexity of DenseNet is similar to FFNN, as residual connections do not add to the network's time complexity. Therefore, time complexity of DenseNet network with $L_D$ layers is in $\mathcal{O}(\sum\nolimits_{l=1}^{L_D} m_lx_l)$, where $m_l$ and $x_l$ denote, respectively, the neuron size and the input size of layer $l$. Assuming an average neuron size equal to the input size of the network (which is the output size of the LSTM network, denoted as $h$), the time complexity of DenseNet can be simplified to $\mathcal{O}(L_Dh^2)$. Finally, the total time complexity of LSTM-DenseNet network in the forward direction can be calculated as follows:
\begin{equation}
    \mathcal{O}_{Tot}=\mathcal{O}((l_{in}+l_p)*(L_{lstm}hN+L_Dh^2)),
\end{equation}
where $l_{in}$ is the input sequence size, $l_p$ is the size of the predicted sequence.

At each step of the rolling algorithm, we feed in the historical data with a size of $l_{in}=l_{buff}$, where $l_{buff}$ is the size of buffer. Subsequently, we predict $l_p$ steps, but only the last $l_f$ steps are retained and appended to the output sequence. Consequently, the time complexity for predicting a single time slot is expressed as follows:
\begin{equation}
    \mathcal{O}_{Roll}=\mathcal{O}(\frac{l_{buff}+l_p}{l_f}*(L_{lstm}hN+L_Dh^2)).
\end{equation}

In the proposed algorithm, at each step, we only feed in the fresh data of size $l_{in}=l_f$, leading to the following time complexity:
\begin{equation}
    \mathcal{O}_{FLSP}=\mathcal{O}(\frac{l_{f}+l_p}{l_f}*(L_{lstm}hN+L_Dh^2)).
\end{equation}
Additionally, it is also necessary to compare the complexity of the proposed algorithm with the traditional rolling algorithm. The proportional complexity of the proposed algorithm to the traditional rolling algorithm can be expressed as follows:
\begin{equation}
    \mathcal{C}=\frac{\mathcal{O}_{FLSP}}{\mathcal{O}_{Roll}}=\mathcal{O}((l_{f}+l_p)/(l_{buff}+l_p)).
\end{equation}
It is worth noting that the proposed algorithm exhibits potential advantages over the traditional rolling algorithm due to its reduced time complexity, especially in scenarios where $l_f$ is considerably smaller than $l_{buff}$. This efficiency improvement becomes particularly valuable for time-sensitive and resource-constrained applications, where rapid and accurate predictions are of paramount importance.

\subsection{Floating-point Operations and Number of Parameters}
In this section, we provide the floating point operations (FLOPs) required by the FLSP and the rolling algorithm to generate the output data. For comparative analysis, we also calculate the FLOPS required by two additional models, the gated recurrent unit (GRU) and the One-dimensional convolutional neural network (CNN-1D). Both GRU and LSTM models, as variants of RNNs, are adept at processing sequential data. However, the GRU architecture is simpler, featuring only two gates. CNN-1D models, applicable to time-series prediction problems, focus primarily on extracting local dependencies within the input data. Unlike RNN-based models that capture broader temporal patterns, CNN-1D models excel in identifying short-term correlations. A limitation of CNN-1D models is that they are only compatible with the rolling algorithm. 
Additionally, the number of parameters for each model is provided, reflecting their training complexity. The FLOPs required by the LSTM-DenseNet network when employing the rolling and FLSP algorithms are detailed as follows:
\begin{equation}
\begin{split}
    FLOP_{Roll}^{(LSTM)}=&\frac{l_{buff}+l_p}{l_f}*(\sum_{l=1}^{L_{LSTM}}(8h_l(h_{l-1}+h_l)\\
    &+29h_l)+\sum_{l=1}^{L_D}(2m_lm_{l-1})),
\end{split}
\end{equation}
\begin{equation}
\begin{split}
    FLOP_{FLSP}^{(LSTM)}=&\frac{l_{f}+l_p}{l_f}*(\sum_{l=1}^{L_{LSTM}}(8h_l(h_{l-1}+h_l)\\
    &+29h_l)+\sum_{l=1}^{L_D}(2m_lm_{l-1})),
\end{split}
\end{equation}
and the number of the model's parameters is:
\begin{equation}
\begin{split}
    P_{LSTM}=\sum_{l=1}^{L_{LSTM}}4h_l(h_{l-1}+h_l+2)+\sum_{l=1}^{L_D}(m_l(m_{l-1}+1))
\end{split}
\end{equation}
where, $h_l$ represents the hidden size of layer $l$, $h_0$ is the input feature size, $L_{LSTM}$ denotes the number of LSTM layers, $L_D$ indicates the number of linear layers in the DenseNet network, $m_{l-1}$ and $m_l$ respectively correspond to the input and output data sizes of layer $l$ in the DenseNet network.
Following the methodology used for the LSTM-DenseNet model, we derive the FLOP calculations and parameter count for the GRU-DenseNet model as follows:
\begin{equation}
\begin{split}
    FLOP_{Roll}^{(GRU)}=&\frac{l_{buff}+l_p}{l_f}*(\sum_{l=1}^{L_{GRU}}(6h_l(h_{l-1}+h_l)+22h)\\
    &+\sum_{l=1}^{L_D}(2m_lm_{l-1})),
\end{split}
\end{equation}
\begin{equation}
\begin{split}
    FLOP_{FLSP}^{(GRU)}=&\frac{l_{f}+l_p}{l_f}*(\sum_{l=1}^{L_{GRU}}(6h_l(h_{l-1}+h_l)+22h)\\
    &+\sum_{l=1}^{L_D}(2m_lm_{l-1})),
\end{split}
\end{equation}
\begin{equation}
    P_{GRU}=\sum_{l=1}^{L_{GRU}}3h_l(h_{l-1}+h_l+2)+\sum_{l=1}^{L_D}(m_l(m_{l-1}+1)),
\end{equation}
where $L_{GRU}$ indicates the number of GRU layers. The FLOP calculations and parameter count for the CNN-1D model can be expressed as follows:
\begin{equation}
\begin{split}
    FLOP&_{Roll}^{(CNN-1D)}=\frac{1}{l_f}*(\sum_{l=1}^{L_{CNN-1D}}(\frac{W}{2^l}C_l(2C_{l-1}K_l\\
    &+1/2))+\sum_{l=1}^{L_{FC}}(2m_lm_{l-1}+m_l)-m_{L_{FC}}),
\end{split}
\end{equation}
\begin{equation}
\begin{split}
    P_{CNN-1D}=&\sum_{l=1}^{L_{CNN-1D}}(C_l(C_{l-1}K+1))\\
    &+\sum_{l=1}^{L_{FC}}(m_l(m_{l-1}+1)),
\end{split}
\end{equation}

where $L_{CNN-1D}$ represents the number of CNN-1D cells, $W$ is the input data size (window size), $C_{l-1}$ and $C_l$ are, respectively, the input and output feature sizes of each CNN-1D cell, $K_l$ denotes the kernel size, $L_{FC}$ indicates the number of fully connected layers, and $m_{l-1}$ and $m_l$ are, respectively, the input and output data sizes of the layer $l$ in the fully connected network.

\subsection{Training the Burst Detection Network}
\label{Burst Detection}
The burst detection network is designed to predict the occurrence of bursty congestion in the following $l_p$ slots. This network gets output chunks of data from the traffic prediction network and determines whether the predicted traffic samples belong to a bursty event. A chunk of data is defined as the concatenation of fresh data (of size $l_f$) and predicted samples (of size $l_p$) at each step. An exemplary chunk of data is shown by a yellow dashed frame in Fig.~\ref{figure5new}(b). The chunks of data are preferred over the output sequence ($\widehat{y}$) as they are smoother and more accurate than the traffic sequence. The samples in the output sequence of the traffic prediction network ($\widehat{y}$) are at least $l_p-l_f$ and at most $l_p$ steps away from the last sample of the real traffic (see Fig.~\ref{figure5new}(b)). Therefor, the input to the burst detection network has a feature size of $ch\_size=2\times(l_f+l_p)$, where the factor of $2$ accounts for the two feature in the output of the traffic prediction network: successfully detected preambles and congested preambles.
According to~\cite{[ref33]}, the prediction accuracy decreases as the number of steps increases. Thus, we prefer to use the data chunks where the predicted points are at least one step and at most $l_p$ steps away from the most recent samples of the real traffic, providing a smaller average lead time from the real data. Moreover, generating data chunks in a single round results in a smoother pattern compared to the output sequence ($\widehat{y}$), which is composed of multiple data segments. This approach aligns with the recommendation of researchers in~\cite{[ref23],[ref57]} who suggest using smoother input sequences in machine learning problems to enhance learning efficiency. Simulations on the validation data show that when $T_{pred}=2\ [sec]$ ($l_p=400$ time slots), using chunks of data rather than predicted streams results in a $76\%$ improvement, confirming our statement. 

\section{Simulation Results}
\label{Sim}
The simulation setup is described in this section, followed by evaluation metrics, simulation results and discussion. The simulation codes used in this study are available at Github\footnote{\url{https://github.com/HosseinMehriB/mMTC-Traffic-Prediction}} for reproducibility and further research.
\subsection{Simulation Setup}
Each base station can support up to a million devices in mMTC networks. Without loss of generality, we consider a single cell with $L=10$ groups of devices and a total number of $N=60,000$ devices. As we are working with the total number of successfully detected and congested preambles, the size of the network and the number of devices does not affect the overall performance of the proposed method. For simplicity, $p_u$ for the periodic data is assumed to be the same for all groups and equal to one packet per $60$ seconds. The probability of occurrence of an event causing bursty traffic in the network is assumed to be different for each group of devices, as illustrated in Table~\ref{table1}. These values are chosen in such a way that the simulation window includes a few bursty events. The duration of each event is assumed a random value between $8$ and $15$ seconds.

\begin{table*}[!t]
\caption{IoT Group Sizes and Event Probability}
\begin{center}
\begin{tabular}{|c|c|c|c|c|c|c|c|c|c|c|}
\hline
\textbf{Group}&\textbf{$\pmb{g_1}$}&\textbf{$\pmb{g_2}$}&\textbf{$\pmb{g_3}$}&\textbf{$\pmb{g_4}$}&\textbf{$\pmb{g_5}$}&\textbf{$\pmb{g_6}$}&\textbf{$\pmb{g_7}$}&\textbf{$\pmb{g_8}$}&\textbf{$\pmb{g_9}$}&\textbf{$\pmb{g_{10}}$}\\
\hline
{\textbf{Size}}&{15000}&{8000}&{3000}&{3000}&{3000}&{15000}&{8000}&{3000}&{2000}&{2000}\\
\hline
{\textbf{Event Probability}}&{0.006}&{0.009}&{0.09}&{0.1}&{0.2}&{0.004}&{0.004}&{0.05}&{0.1}&{0.2}\\
\hline
\end{tabular}
\label{table1}
\end{center}
\end{table*}

Table~\ref{table2} summarizes the RACH configurations that were considered in our simulations~\cite{[ref28]}. We assumed in the simulations that we have $64$ preambles, $10$ of which are reserved, and only $54$ preambles are available. Additionally, it is assumed that the network resources are not shared with other services (such as human-based traffic) and all preambles can be used by IoT devices defined above. PRACH configuration index indicates which subframes of a frame are assigned to the RACH channel. Due to~\cite{[ref28]}, this parameter is determined to be equal to $6$ which means that two subframes in each frame is assigned to the RACH channel. This means that devices try to send their packets over the RACH channel every $0.005$ seconds. Backoff timer determines how long a device should wait before retransmitting a packet after a failure. Moreover, a bursty traffic is labeled as congested when the number of attempts at each time slot exceeds the total available preambles for 250 consecutive time slots.
\begin{table}[htbp]
\caption{RACH Channel Configuration}
\begin{center}
\begin{tabular}{|c|c|}
\hline
\textbf{RACH parameter}&\textbf{Value}\\
\hline
{Number of contention-based preambles}&{54}\\
\hline
{PRACH configuration index}&{6}\\
\hline
{Maximum number of transmissions}&{10}\\
\hline
{Backoff value}&{20}\\
\hline
\end{tabular}
\label{table2}
\end{center}
\end{table}

\begin{table}[htbp]
\caption{Summary of the Simulation Configurations.}
\begin{center}
\begin{tabular}{|c|c|}
\hline
\textbf{ML network parameter}&\textbf{Value}\\
\hline
{$T_{Hist}$ ($l_{Hist}$)}&{$30\ [sec]$ ($6000$ time slots)}\\
\hline
{$T_{fresh}$ ($l_f$)}&{$0.5\ [sec]$ ($100$ time slots)}\\
\hline
{$T_{pred}$ ($l_p$)}&{$\{1,2\} [sec]$ (\{$200$, $400$\} time slots)}\\
\hline
{Total simulation time}&{$160\ [sec]$ ($32000$ time slots)}\\
\hline
{LSTM hidden layer size}&{$2500$}\\
\hline
{FFNN input size}&{$2500$}\\
\hline
{FFNN hidden-1 size}&{$2500$}\\
\hline
{FFNN hidden-2 size}&{$5000$}\\
\hline
{FFNN output size}&{$2$}\\
\hline
{Dropout Value}&{$0.4$}\\
\hline
\end{tabular}
\label{table3}
\end{center}
\end{table}

The input sequences are current and previous values of successfully detected preambles and congested preambles, which are fed to the network as batches of size $25$. The outputs of the proposed network are two sequences: successfully transmitted preambles and congested preambles. Training of the network is similar to the traditional methods. The Adam algorithm is used for the optimization function and mean square error (MSE) as the cost function. During the evaluation phase, an initialization historical data with size of $T_{Hist}=30 [sec]$ ($l_{Hist}=6000$ time slots) is fed to the network and it then starts to predict the output sequences for the next $T_{pred}$ seconds. The fresh data is collected every $0.5$ second and fed to the network. In these simulations, the output sequences are calculated for $T_{pred}=1 [sec]$ ($l_p=200$ time slots) and $T_{pred}=2 [sec]$ ($l_p=400$ time slots). The simulation configurations are summarized in table~\ref{table3}.

\subsection{Data Labeling for Burst Detection Network}
\begin{figure}[htbp]
    \centerline{\includegraphics[scale=0.5]{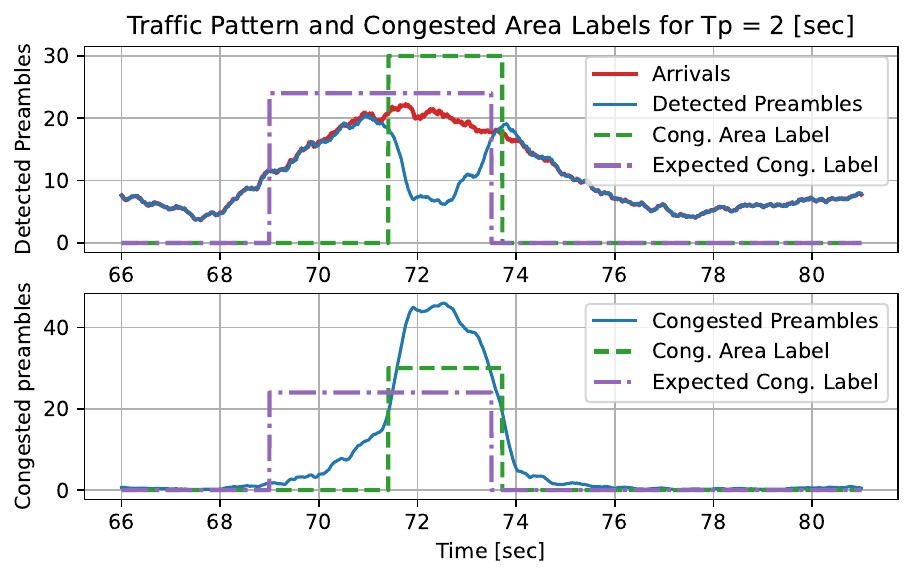}}
    \caption{An example of labeling the congested area for $T_{pred}=2$ seconds. The extracted congestion labels (green dashed line) accurately delineate the congested region where there is a gap between the arrivals and successfully detected preambles.}
    \label{figure60}
\end{figure}

To label the data for the burst detection network, we adopted the definition provided in Subsection~\ref{SysModel}.\ref{Defs}, where an area is labeled as congested if the number of detected attempts is lower than the number of arrivals within a specified time interval. In the simulations, the average number of collided attempts (calculated as the difference between total attempts and successfully detected preambles) was measured over a sliding window of approximately $3$ seconds. This value was then compared to an experimentally determined threshold of $7000$ to identify congested areas. The labeling process, based on these experimental values, demonstrated high accuracy in identifying congested regions. Then, for each prediction time $T_{pred}$, the model is expected to detect congestion $T_{pred}$ seconds prior to its start. Accordingly, the expected congestion labels for the given $T_{pred}$ were assigned a value of ``$1$'' if congestion occurred within the next $T_{pred}$ seconds, and ``$0$'' otherwise. Fig.~\ref{figure60} shows an example of the actual congested area and its corresponding expected congestion labels for $T_{pred}=2$ seconds.

\subsection{Evaluation Metrics}
In the training phase, MSE is used as the cost function for both traffic prediction and burst event detection networks. In the bursty event prediction problem, we usually deal with a kind of imbalanced data consisting of two classes which one of them is a rare event (bursty event samples). In this condition, using MSE metric to evaluate the performance of the burst detection network cannot clearly demonstrate the accuracy of the predictions~\cite{[ref60]}. For example, in a case that $94\%$ of the data is ``$0$'' and only $6\%$ is ``$1$'', a network which always returns ``$0$'' will show an accuracy of $94\%$ in terms of MSE error, but in fact has a terrible performance. On the other hand, a network that detects some of the ``$1$''s correctly and achieves an MSE value of $92\%$, may provide more information than the first network. Thus, besides the MSE metric, three other metrics are used in this paper to show the performance of the burst detection network. The first metric is called precision and defined as~\cite{[ref60]}:
\begin{equation}
\label{eq9}
    Precision = \frac{True\ positives}{True\ positives + False\ positives},
\end{equation}
where, `$True\ positives$' shows the bursty samples that are correctly detected as bursty sample, and `$False\ positives$' are non-bursty samples that incorrectly detected as bursty samples. The second metric is recall and is calculated as~\cite{[ref60]}:
\begin{equation}
\label{eq10}
    Recall = \frac{True\ positives}{True\ positives + False\ negatives},
\end{equation}
where, `$False\ negatives$' are bursty samples that incorrectly detected as non-bursty samples. The denominator of equation~\ref{eq10} includes all detected and missed bursty events which equals to the total number of bursty samples in the target data. The last metric is a combination of two above metrics, called F1-score, and is written as~\cite{[ref60]}:
\begin{equation}
\label{eq11}
    F1\mbox{-}score = \frac{2 \times Precision \times Recall}{Precision + Recall}.
\end{equation}
Among the several existing metrics for rare event detection problems, the mentioned metrics are chosen as they can provide a clear information about the performance of the network. Precision shows how many of the bursty samples in the output of the burst detection network are really a bursty samples. Recall shows how many of the bursty samples are successfully detected by the network. Finally, F1-score which is the combination of precision and recall metrics, shows how accurately the bursty samples are distinguished from the non-bursty samples. We state that a network has good performance when the value of these metrics approaches one.

\subsection{Simulation Results of Traffic Prediction Network}
\begin{figure}[htbp]
\centerline{\includegraphics[scale=0.55]{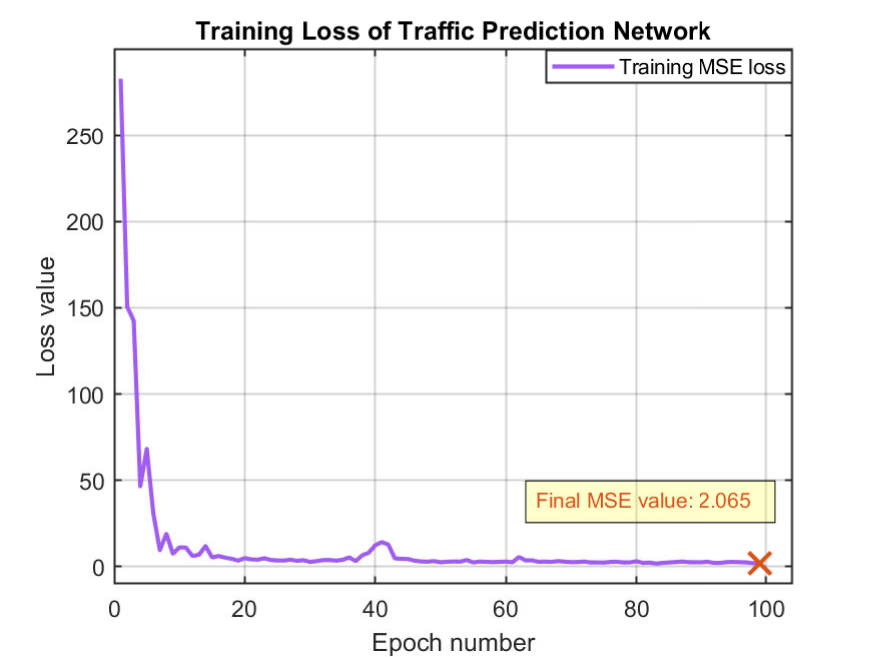}}
\caption{Training loss of traffic prediction network. It reaches to its final value in less than thirty iterations.}
\label{figure5}  
\end{figure}

\begin{figure}[htbp]
\centerline{\includegraphics[scale=0.45]{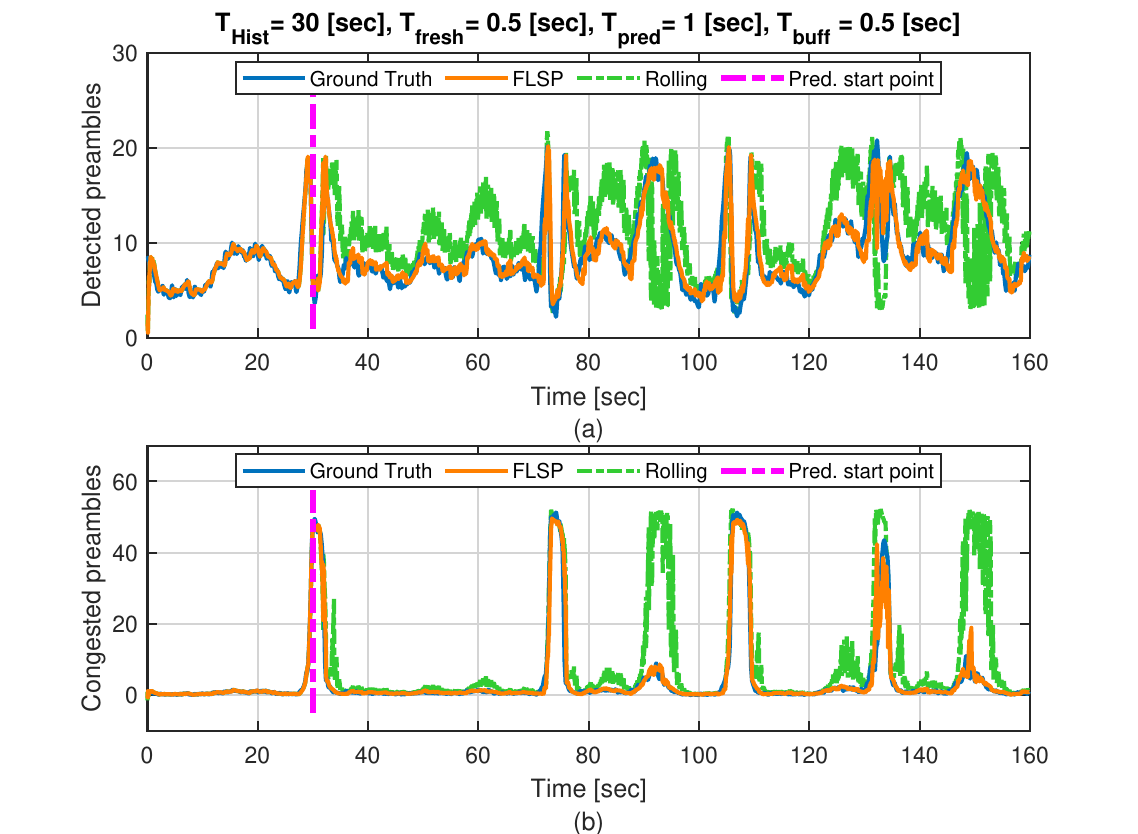}}
\caption{Comparison of predicted traffic patterns and congested preambles using the FLSP and the traditional rolling algorithms for $T_{pred}=1$ [sec] ($l_p=200$ [time slots]). In the rolling method, $0.5$ seconds ($l_{buff}=100$ [time slots]) of historical data is buffered and appended to the fresh data to create the input sequence for the LSTM network. The FLSP algorithm achieves an MSE of $3.96$ compared to $67.73$ in the rolling method, indicating a $94.15\%$ improvement in accuracy. Furthermore, the FLSP algorithm requires $8.05\%$ less time to generate predictions.}
\label{figure6}
\end{figure}

\begin{figure}[htbp]
\centerline{\includegraphics[scale=0.45]{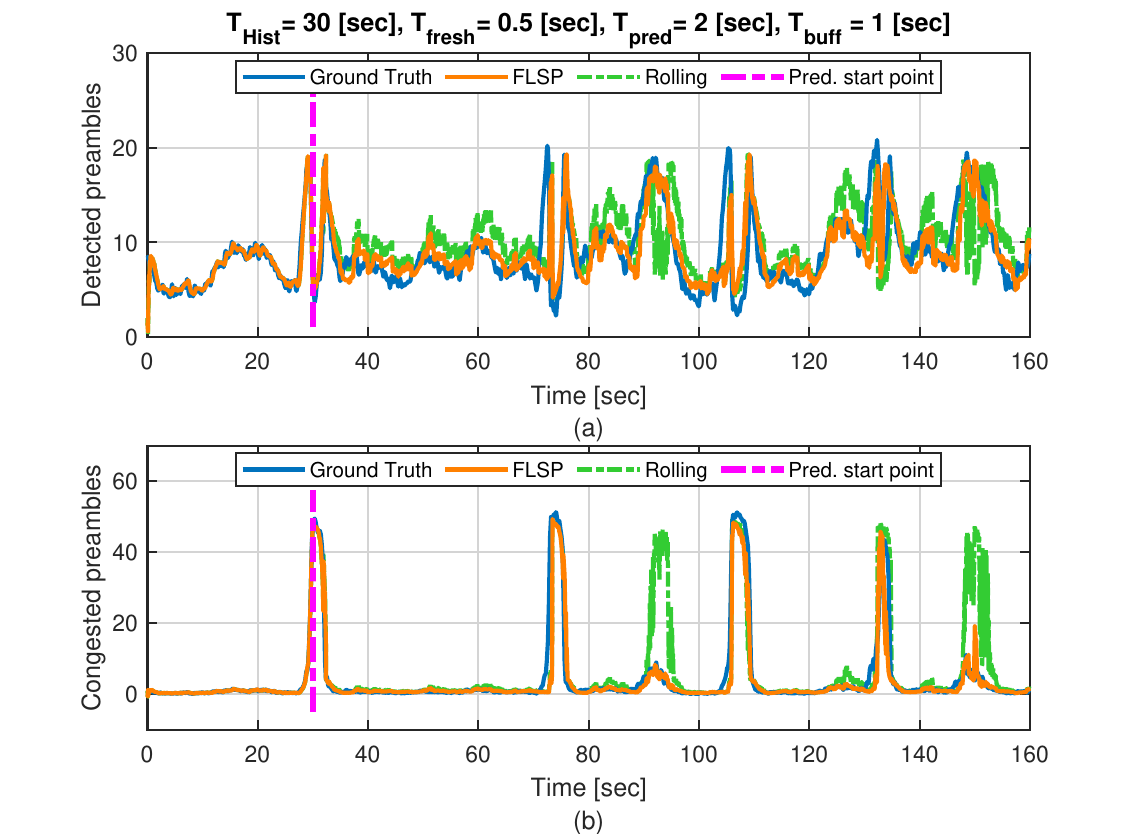}}
\caption{Comparison of predicted traffic patterns and congested preambles using the FLSP and the traditional rolling algorithms for $T_{pred}=2$ [sec] ($l-p=400$ [time slots]). In the rolling method, $1$ second ($l_{buff}=200$ [time slots]) of historical data is buffered and appended to the fresh data to create the input sequence for the LSTM network. The FLSP algorithm achieves an MSE of $11.76$ compared to $39.46$ in the rolling algorithm, indicating a $70.19\%$ improvement in accuracy. Additionally, the FLSP algorithm requires $18.19\%$ less time to generate predictions.}
\label{figure7}
\end{figure}

The training loss of the traffic prediction network is illustrated in Fig.~\ref{figure5}, indicating the network's convergence to a final value in less than thirty iterations. Figures~\ref{figure6} and~\ref{figure7} present the output patterns for $T_{pred}=1 [sec]$ and $T_{pred}=2 [sec]$, respectively, with buffer sizes of $100$ and $200$ time slots for the rolling method. In Fig.~\ref{figure6}, the predicted streams by the FLSP algorithm almost perfectly match the actual streams. The evaluation loss for this configuration is $3.96$ in terms of MSE, representing a substantial improvement of $94.15\%$ over the traditional rolling method, which yields a loss value of $67.73$. Meanwhile, the predictions in Fig.~\ref{figure7}, as expected, demonstrate slightly lower accuracy due to increased prediction time ($T_p$), but still demonstrate an acceptable level of performance. The evaluation loss for this configuration is $11.76$ for FLSP algorithm, signifying a $70.19\%$ improvement compared to the traditional rolling method.

The comparison of the traditional rolling and FLSP algorithm for simulations over a test set of size $100$ sequences is presented in Table~\ref{table9}. Simulation results indicates the proposed method outperforms the traditional method in terms of both accuracy and simulation time by a large margin. As the buffer size increases, the performance of the rolling method approaches that of the FLSP algorithm, eventually reaching the same accuracy when the buffer size is set to $300$ time slots. At this condition, the required time to generate the results using the FLSP algorithm is $23.56\%$ less than that of the rolling method. The buffer size at which two methods achieve the same accuracy largely depends on the properties of the sequential data, where longer dependencies between data points requires a larger buffer size.

\begin{table*}[!htbp]
\centering
\addtolength{\tabcolsep}{-1pt}
\caption{Performance comparison of rolling method and FLSP algorithm when using LSTM for prediction. Input sequence in all cases is concatenation of buffered data and fresh data collected from the mMTC network.}
\label{table9}
\begin{tabular}{*{7}{|c}|}
\hline
\multirow{2}{*}{\textbf{Prediction Method}} & \multirow{2}{*}{\specialcellT{\textbf{Buffer Size}\\ {[time slots] ([sec])}}} & \multirow{2}{*}{\specialcellT{\textbf{Simulation Time}\\\textbf{(change ratio)}}}& \multicolumn{4}{c|}{\textbf{Prediction MSE Value (change ratio)}}\\
\cline{4-7}
& & & $T_p=0.5 [sec]$&$T_p=1 [sec]$&$T_p=1.5 [sec]$&$T_p=2 [sec]$\\
\hline
\specialcellT{FLSP (LSTM)} & NA & $0:41:54$ $(-)$ & $0.44$ $(-)$&$1.61$ $(-)$&$3.52$ $(-)$&$6.78$ $(-)$\\
\hline
\multirow{5}{*}{\specialcellT{{Rolling method}\\{(LSTM)}}} & $0 (0 [sec])$& $0:41:50$ $(-0.16\%)$&$3.38$ $(+86.86\%)$&$33.36$ $(+95.18\%)$&$110.15$ $(+96.81\%)$&$207.85$ $(+96.74\%)$\\
\cline{2-7}
& $100 (0.5 [sec])$& $0:45:34$ $(+8.05\%)$ & $7.07$ $(+93.73\%)$&$40.84$ $(+96.06\%)$&$90.27$ $(+96.10\%)$&$145.89$ $(+95.35\%)$\\
\cline{2-7}
& $200 (1 [sec])$& $0:51:13$ $(+18.19\%)$ & $2.52$ $(+82.40\%)$&$11.31$ $(+85.77\%)$&$18.79$ $(+81.27\%)$&$22.40$ $(+69.71\%)$\\
\cline{2-7}
& $300 (1.5 [sec])$& $0:54:49$ $(+23.56\%)$ & $0.61$ $(+26.82\%)$&$2.14$ $(+24.80\%)$&$3.92$ $(+10.27\%)$&$6.61$ $(-2.57\%)$\\
\cline{2-7}
& $400 (2 [sec])$& $0:58:26$ $(+28.29\%)$ & $0.44$ $(-0.27\%)$&$1.62$ $(+0.38\%)$&$3.53$ $(+0.36\%)$&$6.87$ $(+1.27\%)$\\
\hline
\end{tabular}
\end{table*}

\subsection{Simulation Results of Burst Detection Network}
\begin{figure}[htbp]
\centerline{\includegraphics[scale=0.45]{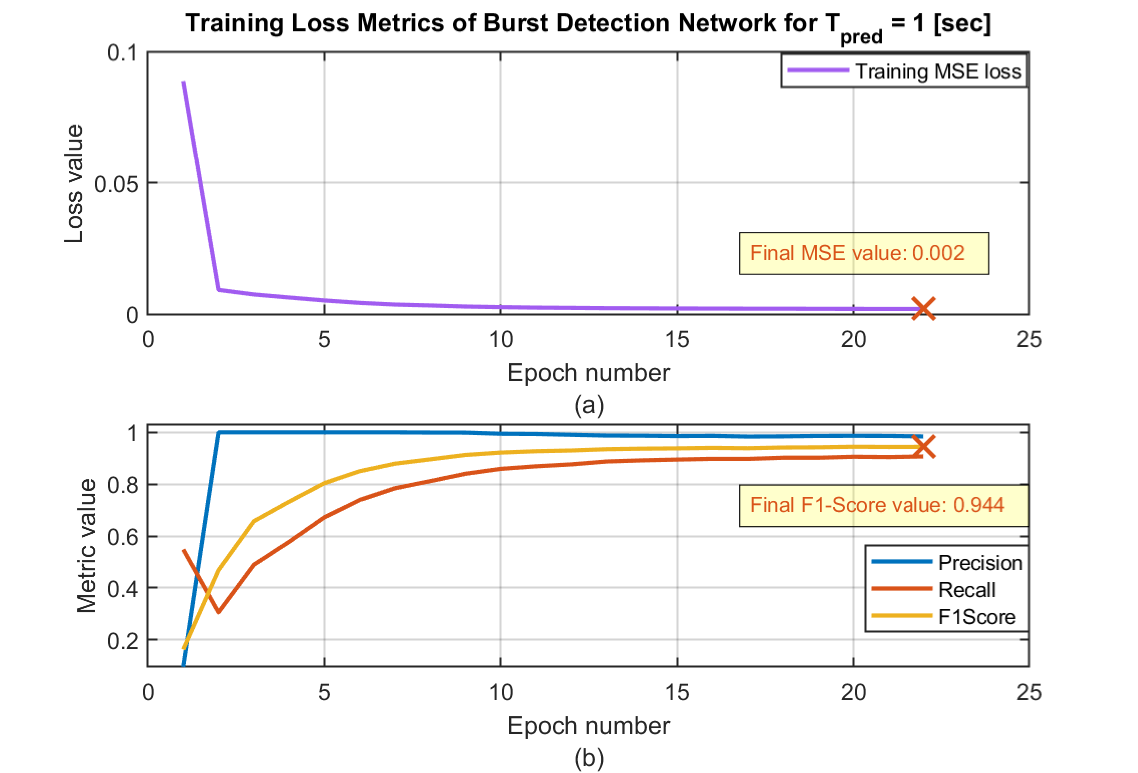}}
\caption{Training loss and metrics for $T_{pred}=1$ [sec]. MSE error remains relatively constant after $10$ epochs, while the F1-Score and recall metrics continue to improve until epoch $15$.}
\label{figure8}
\end{figure}

Performance metrics of the training phase for $T_{pred}=1\ [sec]$ are depicted in Fig.~\ref{figure8}. It is observable from Fig.~\ref{figure8} (a) that the MSE error remains fairly constant after $10$ epochs, while the F1-score and other new metrics improve till epoch $15$ and then stabilize. \ref{table5} summarizes the performance evaluation of burst detection network on the test data sequences. In these simulations, the output of the FLSP and rolling algorithms is fed to the burst detection network as bunches of data. The F1-score metric best illustrates the superiority of the FLSP algorithm in providing high-quality information to the burst detection network, resulting in an F1-score value of $0.92$ out of $1$ for $T_{pred}=1\ [sec]$ and $0.86$ for $T_{pred}=2\ [sec]$.

\begin{table}[htbp]
\hspace{-1pt}
\caption{Validation metrics of burst detection network using outputs of Rolling and FLSP algorithms.}
    \begin{center}
    \begin{tabular}{|c|c|c|c|c|}
    \hline
    \textbf{$\pmb{T_p}$}&\textbf{Prediction Algorithm}&\textbf{Precision}&\textbf{Recall}&\textbf{F1-score}\\
    \hline
    \multirow{3}{*}{\specialcellT{\\\\{$\pmb{1\ [sec]}$}}}&{FLSP}&{$0.95$}&{$0.89$}&{$0.92$}\\
    \cline{2-5}
    &{\specialcellT{{Rolling algorithm}\\{($l_{buff}=100$}\\ {[time slots])}}}&{$0.097$}&{$1$}&{$0.175$}\\
    \cline{2-5}
    &{\specialcellT{{Rolling algorithm}\\{($l_{buff}=200$}\\ {[time slots])}}}&{$0.257$}&{$1$}&{$0.4$}\\
    \hline
    \multirow{3}{*}{\specialcellT{\\\\$\pmb{2\ [sec]}$}}&{FLSP}&{$0.90$}&{$0.83$}&{$0.86$}\\
    \cline{2-5}
    &{\specialcellT{{Rolling algorithm}\\{($l_{buff}=100$}\\ {[time slots])}}}&{$0.093$}&{$0.99$}&{$0.167$}\\
    \cline{2-5}
    &{\specialcellT{{Rolling algorithm}\\{($l_{buff}=200$}\\ {[time slots])}}}&{$0.402$}&{$0.96$}&{$0.555$}\\
    \hline
    \end{tabular}
    \label{table5}
    \end{center}
\end{table}

\begin{figure}[htbp]
\centerline{\includegraphics[scale=0.55]{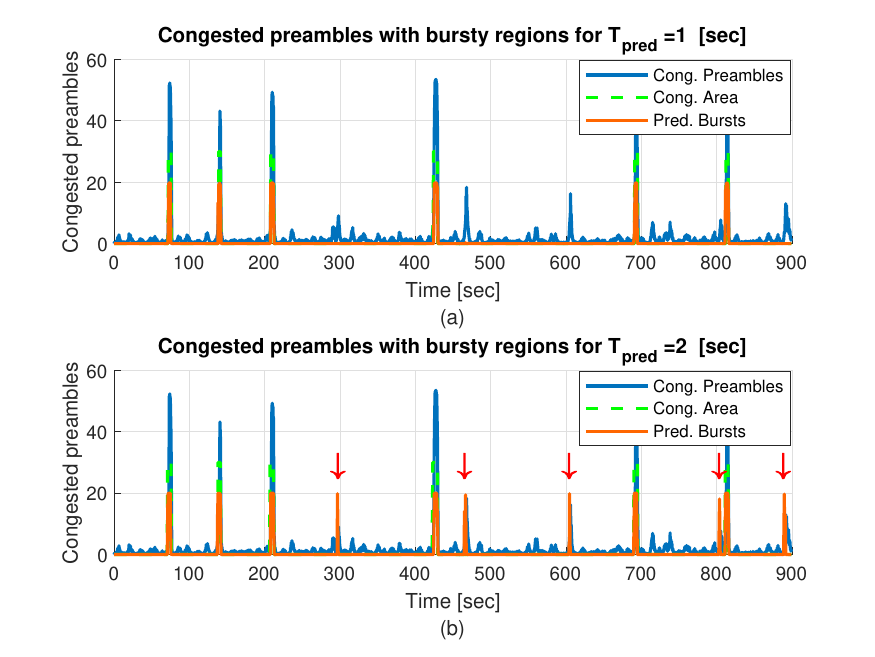}}
\caption{Predicted congested bursty regions (a) for $T_{pred}=1$ [sec], (b) for $T_{pred}=2$ [sec]. The incorrectly detected bursty regions in (b) are indicated by red arrows.}
\label{figure9}
\end{figure}

\begin{figure}[htbp]
\centerline{\includegraphics[scale=0.55]{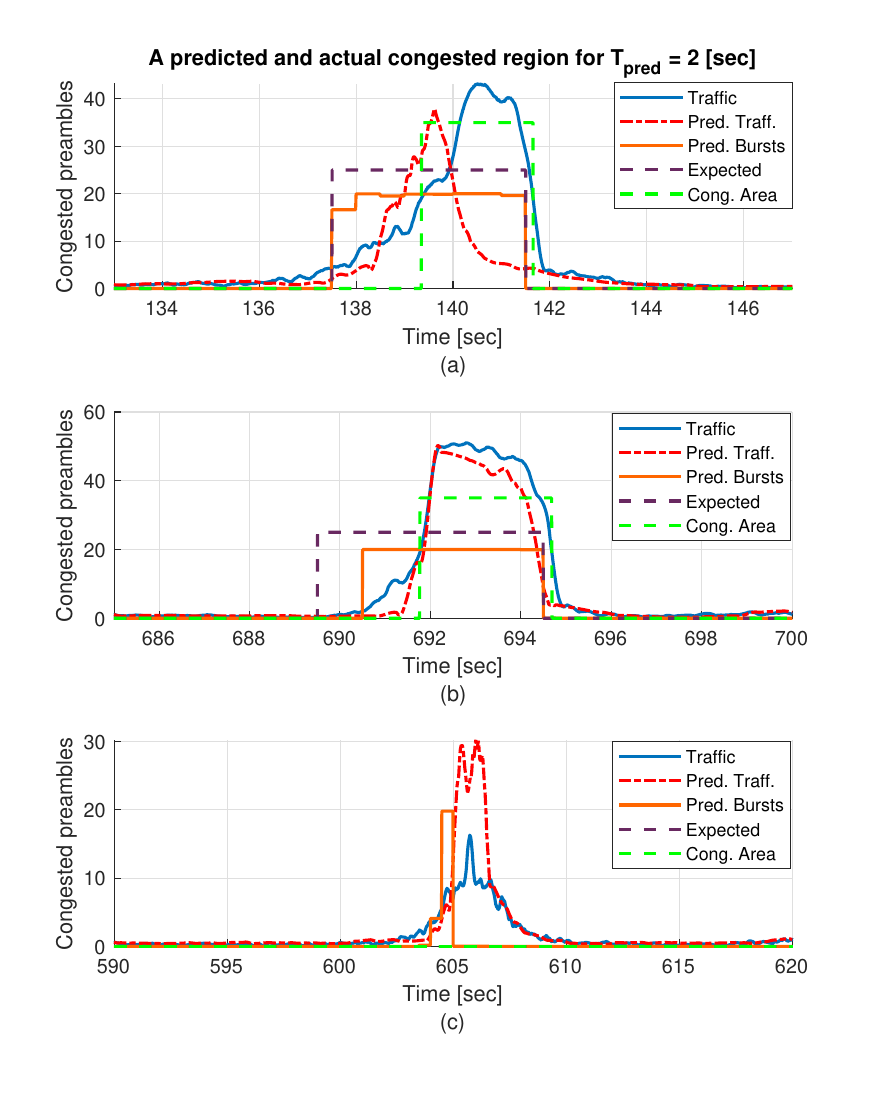}}
\caption{Three examples of the predicted bursty regions for $T_{pred}=2$ [sec]. Figure (a) shows a perfectly predicted bursty region that matches the expected target region. In Figure (b), an example of a late detection of the bursty event is depicted. Figure (c) shows a non-congested region that is predicted as a congested bursty region.}
\label{figure10}
\end{figure}

Fig.~\ref{figure9} shows the predicted bursty regions and actual bursty regions for $T_{pred}=1$ and $2$ seconds. The burst detection network receives the output from the traffic prediction network every $T_{fresh}$ seconds, generating a binary output, either $0$ or $1$, indicating whether congestion is expected within the next $l_f$ time slots. Consequently, for each set of $l_f$ time slots, the burst detection network provides a single value that represents the probability of congestion. For the purposes of visualization in Fig.~\ref{figure9} and~\ref{figure10}, we replicate the burst detection network's output $l_f$ times and plot the scaled results. The scaling is achieved by multiplying the output by $20$ to enhance clarity in the visual representation.
As indicated by the arrows in Fig.~\ref{figure9} (b), some regions are detected as congested by error when $T_{pred}=2 [sec]$, whereas congested regions in~\ref{figure9} (a) are detected almost perfectly when $T_{pred}=1 [sec]$.
Fig.~\ref{figure10} illustrates few examples of perfect and erroneous predictions by the proposed framework. Fig.~\ref{figure10} (a) shows a perfectly predicted bursty region that fits to the expected output. The proposed network predicts the bursty region in Fig.~\ref{figure10} (b) almost a second later than the expected time. Fig.~\ref{figure10} (c) depicts an example of false detection by the proposed network in which a non-congested region is detected as congested region. Late detection and false detection are happen less when $T_{pred}$ is small. As discussed earlier, there is always a trade-off between simulation time, accuracy, and the amount of time available to the network controller. For example, increasing the frequency of fresh data collection (decreasing $l_f$) can increase the resolution of the burst detection outputs and improves the accuracy of the detection at the cost of increased complexity.

\subsection{Comparative Analysis of LSTM, GRU, and CNN-1D Models}

Table~\ref{table10} compares the performance of the LSTM, GRU, and CNN-1D models in predicting mMTC network traffic, along with the number of parameters for each model. The simulation results indicate that the GRU model can achieve performance similar to that of the LSTM model, while requiring fewer parameters and offering faster execution times. For example, the GRU-DenseNet model contains $62,557,502$ parameters and takes $0:23:05$ to generate the output sequence for $100$ streams, whereas the LSTM-DenseNet model has $81,322,502$ parameters and takes $0:51:54$ to generate the same output sequence. Furthermore, applying the FLSP algorithm to the GRU model significantly enhances its performance compared to the rolling algorithm, demonstrating the effectiveness of the proposed FLSP algorithm in improving the GRU model's performance. 

In contrast, the CNN-1D model demonstrates poorer performance due to its inability to capture long-term dependencies. For instance, with a prediction time of $T_{pred}=1$ second and using the FLSP algorithm, LSTM and GRU models achieve MSE of $1.61$ and $2.01$, respectively, while the CNN-1D model records an MSE of $4.20$ with an input size is $200$ time slots. Moreover, when we use the prediction of LSTM and GRU models for detecting the congestion in up coming slots, we achieve a better performance than CNN-1D models. For example, the GRU model achieves an F1-score of $0.92$ with the FLSP algorithm, whereas the CNN-1D model only reaches an F1-score of $0.74$.

Although CNN-1D models are faster than LSTM and GRU models, their performance is inferior, and they lack flexibility. Unlike LSTM and GRU models, which can adapt to varying input sizes and prediction durations, CNN-1D models have fixed input and output sizes, necessitating a new model instance for each configuration. Consequently, LSTM and GRU networks are more favorable for network operators due to their adaptability and higher accuracy.

\begin{table*}[!htbp]
\centering
\addtolength{\tabcolsep}{-1pt}
\caption{Performance comparison of LSTM, GRU, and CNN-1D models.}
\label{table10}
\begin{tabular}{*{8}{|c}|}
\hline
\multirow{2}{*}{\textbf{Prediction Method}} & \multirow{2}{*}{\textbf{Simulation Time}}& \multicolumn{2}{c|}{\textbf{Prediction Accuracy (MSE)}} & \multicolumn{3}{c|}{\textbf{Burst Detection Accuracy (\small{$T_p=1 [sec]$})}} & \multirow{2}{*}{\textbf{Parameter Count}}\\
\cline{3-7}
& & $T_p=1 [sec]$ & $T_p=2 [sec]$ & \textbf{Precision} & \textbf{Recall} & \textbf{F1-score}& \\
\hline
\textbf{LSTM (FLSP)} & $0:41:54$ & $1.61$ & $6.78$ & $0.95$ & $0.89$ & $0.92$ & $81,322,502$\\
\hline
\textbf{GRU (FLSP)} & $0:23:05$ & $1.61$ & $6.30$ & $0.95$ & $0.91$ & $0.93$ & \multirow{3}{*}{\specialcellT{\\ \\ $62,557,502$}} \\
\cline{1-7}
\multirow{2}{*}{\specialcellT{{\textbf{GRU (Rolling)}}\\ {($l_{buff}=100$ [time slots])}}} & \multirow{2}{*}{$0:26:28$} & \multirow{2}{*}{$33.79$} & \multirow{2}{*}{$98.00$} & \multirow{2}{*}{$0.19$} & \multirow{2}{*}{$1$} & \multirow{2}{*}{$0.31$} &  \\
 &  &  &  &  &  &  &  \\
\cline{1-7}
\multirow{2}{*}{\specialcellT{{\textbf{GRU (Rolling)}}\\ {($l_{buff}=200$ [time slots])}}} & \multirow{2}{*}{$0:30:14$} & \multirow{2}{*}{$8.33$} & \multirow{2}{*}{$23.58$} & \multirow{2}{*}{$0.28$} & \multirow{2}{*}{$1$} & \multirow{2}{*}{$0.43$} &  \\
 &  &  &  &  &  &  &  \\
\hline
\multirow{2}{*}{\specialcellT{{\textbf{CNN-1D (Rolling)}}\\ {($W=100$ [time slots])}}} & \multirow{2}{*}{$0:07:19$} & \multirow{2}{*}{$4.87$} & \multirow{2}{*}{$14.76$} & \multirow{2}{*}{$0.71$} & \multirow{2}{*}{$0.78$} & \multirow{2}{*}{$0.74$} & \multirow{2}{*}{$1,267,620$} \\
 &  &  &  &  &  &  &  \\
\cline{1-8}
\multirow{2}{*}{\specialcellT{{\textbf{CNN-1D (Rolling)}}\\ {($W=200$ [time slots])}}} & \multirow{2}{*}{$0:07:18$} & \multirow{2}{*}{$4.20$} & \multirow{2}{*}{$12.11$} & \multirow{2}{*}{$0.75$} & \multirow{2}{*}{$0.82$} & \multirow{2}{*}{$0.78$} & \multirow{2}{*}{$1,657,620$} \\
 &  &  &  &  &  &  &  \\
\hline
\end{tabular}
\end{table*}
\section{Conclusion and Future Work}
\label{Conc}In this paper, we presented a novel ML framework for predicting successfully detected and congested preambles, as well as identifying congested bursty regions in an mMTC network. Leveraging the strengths of LSTM and DenseNet networks along with a new prediction algorithm, enabled our proposed approach to demonstrated superior performance compared to traditional time-series prediction methods. By efficiently storing hidden states instead of historical data, our method retained longer dependencies, effectively mitigating the limitations of the rolling approach with limited buffer size. Complexity studies have indicated that the FLSP algorithm requires only a fraction of the processing time of the traditional rolling algorithm while achieving better accuracy. This makes the proposed method well-suited for time-critical and resource-limited applications. 
Moreover, including the information of both transmitted traffic and congested preambles in the calculations not only improved the accuracy of the predictions, but also provided useful insights into the bursty congested regions. The output of the traffic prediction network is fed into the burst detection network in chunks of data, and the bursty regions are generated as output, predicting the congestion before it occurs.
Simulation results confirm that the predicted values closely match the target values with minimal error. The advantages of our proposed approach can be summarized as follows:
\begin{itemize}
    \item Full utilization of available information: Our method leverages the full potential of available information in mMTC networks, including statistics of congested preambles and frequent fresh data collected from the network.
    \item High accuracy and low processing load: The proposed method achieves high accuracy in real-world environments with complicated patterns without imposing an extra processing load on the network.
    \item Proactive congestion handling: Our networks provide accurate information about bursty regions a few seconds in advance, enabling proactive methods to handle network congestion and reduce packet loss.
    \item Flexible parameter adjustments: The adjustable parameters of our proposed method offer flexibility to designers, allowing them to tailor the predictions to meet specific network requirements in terms of accuracy, processing load, and prediction duration.
    \item Applicability to various ML architectures: The FLSP algorithm is applicable to a wide range of ML architectures where continuous predictions are required.
\end{itemize}
Future work will focus on optimizing the parameters of our method and extracting the equations and metrics for the design problem. To better understand the inner workings of the proposed ML architecture and the effectiveness of each parameter and input feature on network performance, we will employ explainable deep learning techniques as discussed in~\cite{[ref602]}. We will also investigate the effectiveness of the FLSP algorithm when applied to other types of recurrent neural networks, such as ConvLSTM, with the goal of source traffic prediction. Additionally, given the superior ability of our proposed algorithm to predict time-series data accurately, we plan to apply it to new problems requiring precise predictions, such as predicting the age of information (AOI) in time-critical scenarios. By further exploring these avenues, we aim to enhance the performance and applicability of our ML framework and contribute to the advancement of mMTC network capabilities and efficiency.

\newpage

\begin{IEEEbiography}[{\includegraphics[width=1in,height=1.25in,clip,keepaspectratio]{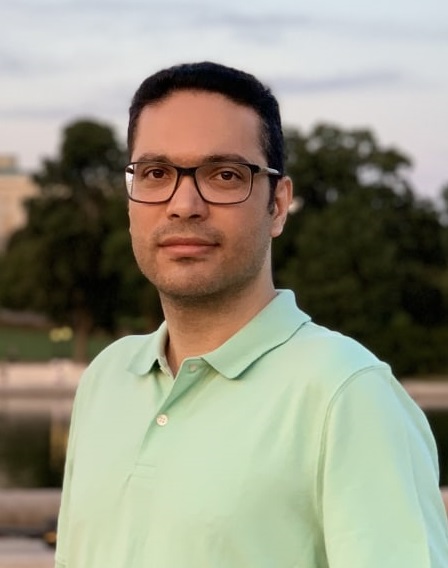}}]
{Hossein Mehri}~(Student member, IEEE)~received the B.Sc. degree from Zanjan National University, Zanjan, Iran, in 2010, the M.Sc. degree in electrical engineering from Iran University of Science and Technology, Tehran, Iran, in 2013, and the Ph.D. degree in electrical and computer engineering from Boise State University, Boise, ID, USA in 2024. He is currently a research scientist at Boise State University, Boise, ID, USA. His current research interests include machine learning, wireless communications, and MAC layer protocols.
\end{IEEEbiography}

\begin{IEEEbiography}
[{\includegraphics[width=1in,height=1.25in,clip,keepaspectratio]{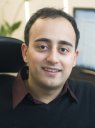}}]
{Hani Mehrpouyan}~(Member, IEEE)~received the B.Sc. degree (Hons.) from Simon Fraser University, Burnaby, Canada, in 2004, and the Ph.D. degree in electrical engineering from Queens University, Kingston, Canada, in 2010. From September 2011 to March 2012, he held a post-doctoral position at the Chalmers University of Technology, where he led the MIMO aspects of the microwave backhauling for next-generation wireless networks project. He was an Assistant Professor at California State University, Bakersfield, CA, USA, from 2012 to 2015, and Boise State University, Boise, ID, USA, from 2015 to 2022. He has received the IEEE Conference on Communication (ICC) Best Paper Award from the Communication Theory Symposium.
\end{IEEEbiography}

\begin{IEEEbiography}
[{\includegraphics[width=1in,height=1.25in,clip,keepaspectratio]{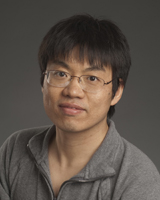}}]
{Hao Chen}~(Member, IEEE)~received the Ph.D. degree in electrical engineering from Syracuse University, Syracuse, NY, USA, in 2007. From 2007 to 2008 and from 2008 to 2010, he was a Postdoctoral Research Associate and Research Assistant Professor with Syracuse University. Since August 2010, he has been an Assistant Professor with the Department of Electrical and Computer Engineering and Computer Science, Boise State University, Boise, ID, USA. His current research interests include statistical signal and image processing, and communications.
\end{IEEEbiography}

\vfill\pagebreak


\begin{thebibliography}{00}
\bibitem{[ref1]} L. Chettri and R. Bera, ``A Comprehensive Survey on Internet of Things (IoT) Toward 5G Wireless Systems," in \textit{IEEE Internet Things J.}, vol. 7, no. 1, pp. 16-32, Jan. 2020.
\bibitem{[ref2]} C. Du, L. Xie, Z. Xing, J. An and Z. Zhang, ``Self-Interference Cancellation Based Workload-Driven Duplex-Model Selection in Machine-type Communications Networks," in \textit{IEEE Internet Things J.}, May 2022.
\bibitem{[ref3]} H. Shariatmadari \textit{et al.}, ``Machine-type communications: current status and future perspectives toward 5g systems," in \textit{IEEE Commun. Mag.}, vol. 53, no. 9, pp. 10–17, Sep. 2015.
\bibitem{[ref4]} J. Navarro-Ortiz, P. Romero-Diaz, S. Sendra, P. Ameigeiras, J. J. Ramos-Munoz and J. M. Lopez-Soler, ``A Survey on 5G Usage Scenarios and Traffic Models," in \textit{IEEE Communications Surveys \& Tutorials}, vol. 22, no. 2, pp. 905-929, Feb. 2020.
\bibitem{[ref604]} S. I. Khan \textit{et al.}, ``Implementation of cloud based IoT technology in manufacturing industry for smart control of manufacturing process,'' in \textit{International Journal on Interactive Design and Manufacturing (IJIDeM)}, pp. 1-13, 2023.
\bibitem{[ref605]} G. Tabella \textit{et al.}, ``Bayesian Fault Detection and Localization Through Wireless Sensor Networks in Industrial Plants,'' in \textit{IEEE Internet of Things J.}, vol. 11, no. 8, pp. 13231-13246, Apr. 2024.
\bibitem{[ref6]} S. K. Sharma and X. Wang, ``Toward Massive Machine Type Communications in Ultra-Dense Cellular IoT Networks: Current Issues and Machine Learning-Assisted Solutions," \textit{IEEE Commun. Surveys Tuts.}, vol. 22, no. 1, pp. 426-471, Firstquarter 2020.
\bibitem{[ref9]} D. Zucchetto and A. Zanella, ``Multi-rate ALOHA Protocols for Machine-Type Communication," in \textit{2018 International Conference on Computing, Networking and Communications (ICNC)}, pp. 524-530, Mar. 2018.
\bibitem{[ref11]} O.A. Amodu, M. Othman, ``A survey of hybrid MAC protocols for machine-to-machine communications," in \textit{Telecommun. Syst.}, vol. 69, pp. 141–165, 2018.
\bibitem{[ref12]} N. Jiang, Y. Deng and A. Nallanathan, ``Traffic Prediction and Random Access Control Optimization: Learning and Non-Learning-Based Approaches," in \textit{IEEE Commun. Mag.}, vol. 59, no. 3, pp. 16-22, Mar. 2021.
\bibitem{[ref13]} M. S. Ali, E. Hossain and D. I. Kim, ``LTE/LTE-A Random Access for Massive Machine-Type Communications in Smart Cities," \textit{IEEE Commun. Mag.}, vol. 55, no. 1, pp. 76-83, Jan. 2017.
\bibitem{[ref10]} A. Laya, C. Kalalas, F. Vazquez-Gallego, L. Alonso and J. Alonso-Zarate, ``Goodbye, ALOHA!," in \textit{IEEE Access}, vol. 4, pp. 2029-2044, Apr. 2016.
\bibitem{[ref17]} T. Hara, K. Ishibashi, ``Grant-Free Non-Orthogonal Multiple Access With Multiple-Antenna Base Station and Its Efficient Receiver Design", in \textit{IEEE Access}, vol. 7, pp. 175717-175726, Nov. 2019.
\bibitem{[ref18]} H. Jiang, D. Qu, J. Ding, T. Jiang, ``Multiple Preambles for High Success Rate of Grant-Free Random Access With Massive MIMO," in \textit{IEEE Trans. Wireless Commun.}, vol. 18, no. 10, pp. 4779-4789, Jul. 2019.
\bibitem{[ref19]} W. Zhou \textit{et al.}, ``A Survey on Transmission Schemes on Large-Scale Internet of Things with Nonorthogonal Multiple Access," in \textit{Wireless Communications and Mobile Computing}, vol. 2021, pp. 1-11, Sep. 2021.
\bibitem{[ref20]} Laner, M., Nikaein, N., Drajic, D., Svoboda, P., Popovic, M. and Krco, S., ``M2M traffic and models," \textit{Machine-to-Machine Communications: Architectures, Technology, Standards, and Applications}, Boca Raton, FL, USA:CRC, 2014.
\bibitem{[ref21]} 3GPP TR 23.888, ``System Improvements for Machine-type Communications,'' R11, V1.3.10, 2012.
\bibitem{[ref22]} L. Ferdouse, A. Anpalagan, and S. Misra, ``Congestion and overload control techniques in massive M2M systems: a survey," in \textit{Trans. on Emerg. Telecommun. Technol.}, vol. 28, no. 2, 2017.
\bibitem{[ref23]} A. S\o{}raa, T. N. Weerasinghe, I. A. M. Balapuwaduge and F. Y. Li, ``Preamble Transmission Prediction for mMTC Bursty Traffic: A Machine Learning based Approach," in \textit{IEEE Global Commun. Conf.}, pp. 1-6, 2020.
\bibitem{[ref24]} 3GPP TS 36.321, ``Evolved Universal Terrestrial Radio Access (E-UTRA), Medium Access Control (MAC) Protocol Specification,” R16, V16.0.0, Mar. 2020.
\bibitem{[ref25]} A. D. Shoaei, D. T. Nguyen and T. Le-Ngoc, ``Traffic Prediction for Reconfigurable Access Scheme in Correlated Traffic MTC Networks," in \textit{IEEE 32nd Ann. Int. Symp. On Personal, Indoor and Mobile Radio Commun. (PIMRC)}, pp. 953-958, 2021.
\bibitem{[ref26]} Simmhan, Y., Ravindra, P., Chaturvedi, S., Hegde, M. and Ballamajalu, R., ``Towards a Data‐driven IoT Software Architecture for Smart City Utilities," in \textit{Softw.: Pract. and Experience}, vol. 48, no. 7, pp.1390-1416, 2018.
\bibitem{[ref27]} V. Rodoplu, M. Nakıp, D. T. Eliiyi and C. Güzeliş, ``A Multiscale Algorithm for Joint Forecasting–Scheduling to Solve the Massive Access Problem of IoT," in \textit{IEEE Internet Things J.}, vol. 7, no. 9, pp. 8572-8589, Sep. 2020.
\bibitem{[ref28]} 3GPP TR 37.868, ``Study on RAN Improvements for Machine-type Communications", R11, V11.0.0, Sep. 2011.
\bibitem{[ref29]} L. Liu, D. Essam and T. Lynar, ``Complexity Measures for IoT Network Traffic," in \textit{IEEE Internet Things J.}, early access, 2022, doi: 10.1109/JIOT.2022.3197323.
\bibitem{[ref30]} G. E. P. Box \textit{et al.}, ``Model Diagnostic Checking," in \textit{Time Series Analysis: Forecasting and Control}, 5th ed., Hoboken, NJ, USA: Wiley,pp. 284-305, 2016.
\bibitem{[ref31]} S. Ali, W. Saad and N. Rajatheva, ``A Directed Information Learning Framework for Event-Driven M2M Traffic Prediction," \textit{IEEE Commun. Lett.}, vol. 22, no. 11, pp. 2378-2381, Nov. 2018.
\bibitem{[ref32]} H. D. Trinh, L. Giupponi, and P. Dini, ``Mobile traffic prediction from raw data using LSTM networks,” in  \textit{29th IEEE Annu. Int. Symp. on Personal, Indoor and Mobile Radio Commun., PIMRC}, pp. 1827–1832, 2018.
\bibitem{[ref33]} R. Chandra, S. Goyal and R. Gupta, ``Evaluation of Deep Learning Models for Multi-Step Ahead Time Series Prediction," in \textit{IEEE Access}, vol. 9, pp. 83105-83123, 2021.
\bibitem{[ref37]} R. M. French, ``Catastrophic forgetting in connectionist networks," in \textit{Trends Cogn. Sci.}, vol. 3, no. 4, pp. 128-135, Apr. 1999.
\bibitem{[ref38]}M. McCloskey and N. J. Cohen, ``Catastrophic interference in connectionist networks: The sequential learning problem," in \textit{Psychology of learning and motivation}. Elsevier, vol. 24, pp. 109–165, 1989.
\bibitem{[ref39]} C. Wei, G. Bianchi and R. Cheng, ``Modeling and Analysis of Random Access Channels With Bursty Arrivals in OFDMA Wireless Networks," \textit{IEEE Trans. Wireless Commun.}, vol. 14, no. 4, pp. 1940-1953, Apr. 2015.
\bibitem{[ref40]} J. Vidal, L. Tello-Oquendo, V. Pla and L. Guijarro, ``Performance Study and Enhancement of Access Barring for Massive Machine-Type Communications," in \textit{IEEE Access}, vol. 7, pp. 63745-63759, 2019.
\bibitem{[ref41]} J. Park and Y. Lim, ``Random-Access Control Method for MTC in an LTE System Inspired by the Minority Game," in \textit{IEEE Trans. Veh. Technol.}, vol. 67, no. 9, pp. 9037-9041, Sep. 2018.
\bibitem{[ref42]} O. Arouk, A. Ksentini and T. Taleb, ``Group Paging-Based Energy Saving for Massive MTC Accesses in LTE and Beyond Networks," in \textit{IEEE J. Sel. Areas Commun.}, vol. 34, no. 5, pp. 1086-1102, May 2016.
\bibitem{[ref43]} M. Bouzouita, Y. Hadjadj‐Aoul, N. Zangar, and G. Rubino, ``Estimating the number of contending IoT devices in 5G networks: Revealing the invisible," in \textit{Transactions on emerging telecommunications technologies}, vol. 30, no. 4, 2019.
\bibitem{[ref44]} Tello-Oquendo, L., Leyva-Mayorga, I., Pla, V., Martinez-Bauset, J., Vidal, J.R., Casares-Giner, V. and Guijarro, L., ``Performance analysis and optimal access class barring parameter configuration in LTE-A networks with massive M2M traffic," in \textit{IEEE Trans. Veh. Technol.}, vol. 67, no. 4, pp.3505-3520, 2017.
\bibitem{[ref45]}Y. Hadjadj-Aoul and S. Ait-Chellouche, ``Access Control in NB-IoT Networks: A Deep Reinforcement Learning Strategy," \textit{Information}, vol. 11, no. 11, p. 541, Nov. 2020.
\bibitem{[ref49]} T. Senevirathna, B. Thennakoon, T. Sankalpa, C. Seneviratne, S. Ali and N. Rajatheva, ``Event-Driven Source Traffic Prediction in Machine-Type Communications Using LSTM Networks," in \textit{IEEE Global Commun. Conf.}, pp. 1-6, 2020.
\bibitem{[ref50]} E. Eldeeb, M. Shehab and H. Alves, ``A Learning-Based Fast Uplink Grant for Massive IoT via Support Vector Machines and Long Short-Term Memory," in \textit{IEEE Internet Things J.}, vol. 9, no. 5, pp. 3889-3898, Mar. 2022.
\bibitem{[ref51]} M. E. Tanab and W. Hamouda, ``Efficient Resource Allocation in Fast-Uplink Grant for Machine-Type Communications With NOMA," in \textit{IEEE Internet Things J.}, vol. 9, no. 18, pp. 18113-18129, 15 Sep. 2022.
\bibitem{[ref52]} M. Shehab, A. K. Hagelskj\ae{}r, A. E. Kal\o{}r, P. Popovski and H. Alves, ``Traffic Prediction Based Fast Uplink Grant for Massive IoT," in \textit{IEEE 31st Ann. Int. Symp. On Personal, Indoor and Mobile Radio Commun. (PIMRC)}, pp. 1-6, 2020.
\bibitem{[ref53]} Y. Cao, R. Wang, M. Chen and A. Barnawi, ``AI Agent in Software-Defined Network: Agent-Based Network Service Prediction and Wireless Resource Scheduling Optimization," in \textit{IEEE Internet Things J.}, vol. 7, no. 7, pp. 5816-5826, Jul. 2020.
\bibitem{[ref54]} S. K. Sharma, and X. Wang, ``Collaborative Distributed Q-Learning for RACH Congestion Minimization in Cellular IoT Networks," in \textit{IEEE Commun. Lett.}, vol. 23, no. 4, pp. 600-603, Apr. 2019.
\bibitem{[ref55]} D. -D. Tran, S. K. Sharma, and S. Chatzinotas, ``BLER-based Adaptive Q-learning for Efficient Random Access in NOMA-based mMTC Networks," in \textit{IEEE 93rd Veh. Technol. Conf. (VTC2021-Spring)}, pp. 1-5, 2021.
\bibitem{[ref57]} T. N. Weerasinghe, I. A. M. Balapuwaduge and F. Y. Li, ``Supervised Learning based Arrival Prediction and Dynamic Preamble Allocation for Bursty Traffic," in \textit{IEEE Conf. Comput. Commun. Workshops (INFOCOM WKSHPS)}, pp. 1-6, 2019.
\bibitem{[ref58]} A. Laya, L. Alonso, and J. Alonso-Zarate, ``Is the random access channel of LTE and LTE-A suitable for M2M communications? A survey of alternatives," \textit{IEEE Commun. Surveys Tuts.}, vol. 16, no. 1, pp. 4–16, 2014.
\bibitem{[ref59]} K. He, X. Zhang, S. Ren, and J. Sun. ``Deep residual learning for image recognition," in \textit{Proc. Of the IEEE Conf. on Computer Vision and Pattern Recognition}, pp. 770-778, 2016.
\bibitem{[ref59-2]} G. Huang \textit{et al.}, ``Densely Connected Convolutional Networks," in \textit{Proc. Of the IEEE Conf. on Computer Vision and Pattern Recognition},pp. 4700-4708, 2017.
\bibitem{[ref59-3]} C. Zhang \textit{et al.}, ``Resnet or DenseNet? Introducing Dense Shortcuts to Resnet," in \textit{Proc. of the IEEE/CVF Winter Conf. on Appl. of Computer Vision}, pp. 3550-3559, 2021.
\bibitem{[ref60]} P. Branco, L. Torgo, R. P. Ribeiro, ``A Survey of Predictive Modeling on Imbalanced Domains,'' in \textit{ACM Comput. Surv.}, vol. 49, no. 2, pp. 1-50, Nov. 2016.
\bibitem{[ref602]} I. Guarino \textit{et al.}, ``Explainable Deep-Learning Approaches for Packet-Level Traffic Prediction of Collaboration and Communication Mobile Apps,'' in \textit{IEEE Open J. Commun. Soc.}, vol. 5, pp. 1299-1324, Feb. 2024.
\end{thebibliography}
\end{document}